\documentclass[twocolumn,trackchanges,floatfix]{aastex7}

\usepackage{graphicx}
\usepackage{textgreek}
\usepackage{amsmath}
\usepackage{lineno}
\usepackage{csquotes}
\usepackage{gensymb}
\usepackage{subcaption}  
\usepackage{hyperref}
\usepackage{textcomp}

\begin{document}

\title{On the estimation of Sulfuric Acid Vapor concentrations below the Venus cloud deck using the Akatsuki Radio Science Experiment}

\author[orcid=0009-0001-6731-3015]{S. Banerjee}
\affiliation{Space Physics Laboratory, VSSC, Thiruvananthapuram, India}
\affiliation{Research Centre, University of Kerala,  Thiruvananthapuram, 695034, India}
\email[show]{soumyaneal08@gmail.com}  

\author[orcid=0000-0002-1276-0088]{R. K. Choudhary}
\affiliation{Space Physics Laboratory, VSSC, Thiruvananthapuram, India}
\email{rajkumar.choudhary@gmail.com}

\author[orcid=0000-0001-9027-3202]{K. R. Tripathi}
\affiliation{Graduate School of Frontier Sciences, The University of Tokyo, Kashiwa, Japan}
\email{krtripathi95@google.com}

\author[orcid=0000-0002-9470-4492]{T. Imamura}
\affiliation{Graduate School of Frontier Sciences, The University of Tokyo, Kashiwa, Japan}
\email{t_imamura@edu.k.u-tokyo.ac.jp}

\author[orcid=0000-0001-5088-6186]{H. Ando}
\affiliation{Faculty of Science, Kyoto Sangyo University, Kyoto, Japan}
\email{hando@cc.kyoto-su.ac.jp}

\begin{abstract}
We report new constraints on the vertical distribution of sulfuric acid vapor in the Venusian atmosphere, derived from a refined analysis of radio occultation (RO) data. The method estimates the power spectral density (PSD) of the received signal to recover both the signal intensity and the Doppler shift. The received signal power is estimated at 1-sec cadence which enhances the sensitivity and detection of the signal at lower altitudes of Venus, even in regions of high atmospheric opacity. After correcting total attenuation for refractive losses, absorption by known microwave absorbers is removed, leaving a residual signal attributable to sulfuric acid vapor. Two different methods of estimating the absorption due to Sulfur Dioxide have been presented, including one which incorporates in-situ data, which should better constrain the sulfuric acid vapor abundance below the clouds. Retrieved profiles for altitudes of 40 - 50 km reveal an increasing vapor abundance to more than 10 ppm below the clouds, and a sharp decline above 50 km in line with the expected saturation profile. These measurements agree with current models of the Venusian cloud structure and composition, and demonstrate that RO data, when coupled with optimized spectral analysis, can yield quantitative constraints on trace absorbers in optically thick atmospheres.
\end{abstract}


\section{Introduction} 

The Radio Occultation (RO) technique has long been employed to investigate the dense atmosphere of Venus. Initially proposed in the 1960s to derive neutral atmospheric parameters such as temperature and pressure, as well as to study ionospheric perturbations, RO makes use of phase fluctuations in spacecraft telemetry signals as they pass through a planetary atmosphere \citep{https://doi.org/10.1029/JZ070i013p03217}. By the 1970s, it was recognized that the monotonic attenuation of radio signals during ingress and egress could also be used to infer atmospheric composition, provided the signal underwent measurable absorption by atmospheric constituents \citep{fjeldbo1971neutral}. Such absorption, particularly in the microwave region (S- and X-bands), was first detected during the Mariner V flyby of Venus at S-band frequencies (2297 MHz) \citep{fjeldbo1971neutral}. Subsequent Venus missions—including Pioneer Venus, Venera, Magellan, Venus Express, and more recently Akatsuki—have continued to report similar attenuation signatures \citep{cimino1982composition, steffes1981laboratory, gubenko2001radio, steffes1994radio, oschlisniok2012microwave, imamura2017initial}.

Early laboratory studies suggested CO$_2$, N$_2$, and SO$_2$ as possible microwave absorbers under Venus-like conditions. While SO$_2$, due to its permanent dipole moment, was initially believed to be the primary absorber \citep{janssen1981microwave}, observed attenuation levels exceeded the predicted contributions from these species alone \citep{steffes1981laboratory}. Other candidates such as H$_2$O, OCS, NH$_3$, and SO$_3$ were ruled out based on their weak absorptivity and/or negligible abundances \citep{steffes1982sulfuric}.

This discrepancy was resolved when \citet{steffes1982sulfuric} identified gaseous H$_2$SO$_4$ as an exceptionally strong microwave absorber under simulated Venusian conditions. Its weak frequency dependence in the microwave range explained both occultation and ground-based observations that had previously attributed most absorption to SO$_2$ \citep{steffes1981laboratory, janssen1981constraints}. It is now established that sulfuric acid vapor, present primarily in the 35 -50 km altitude range below the main cloud deck, accounts for the dominant microwave opacity in Venus' atmosphere \citep{young1973clouds, hansen1974interpretation}.

Subsequent studies aimed to quantify the H$_2$SO$_4$ vapor abundance. Using refined absorptivity measurements and empirical models, \citet{kolodner1998microwave} derived vertical profiles from Magellan and Mariner 10 RO data, reporting peak abundances of 10 -14 ppm near the equator, tapering to 3 -7 ppm at mid-latitudes, and increasing again at high latitudes. Similar latitudinal patterns were later confirmed using over 800 profiles from Venus Express' VeRa payload \citep{oschlisniok2021sulfuric}. These spatial variations were attributed to meridional circulation patterns akin to Hadley cells, involving upwelling in equatorial regions and subsidence near the poles.

The origin and cycling of H$_2$SO$_4$ in the Venusian atmosphere are known to be closely tied to sulfur chemistry and planetary dynamics. SO$_2$, produced by volcanic outgassing \citep{bullock2001recent}, rises to the upper atmosphere where it reacts with photochemically produced oxygen to form SO$_3$, which then hydrates to form H$_2$SO$_4$ vapor \citep{hashimoto2001predictions}. The vapor condenses into liquid droplets near the $\sim$62 km altitude, forming the characteristic sulfuric acid clouds. These droplets are then transported poleward and downward, re-evaporating around 48 -50 km, and in some cases decomposing into SO$_3$ and H$_2$O below 40 km \citep{ragent1980structure, krasnopolsky1994h2o}. Thus, sulfuric acid vapor serves as a key tracer of atmospheric dynamics and chemical cycling on Venus.

In addition to its role in cloud formation and radiative balance, H$_2$SO$_4$ vapor also provides an indirect means of estimating SO$_2$ abundance in the 51 -54 km range, as shown by \citet{oschlisniok2021sulfuric}. Given that SO$_2$ is considered a proxy for volcanic activity \citep{oyama1979venus}, long-term monitoring of H$_2$SO$_4$ profiles can offer insights into both atmospheric dynamics and possible geologic activity \citep{esposito1997chemistry, na1990international}.

Despite the acknowledged utility of RO in retrieving H$_2$SO$_4$ profiles, the lack of openly available, end-to-end signal processing frameworks remains a barrier. Most existing studies rely on proprietary or specialized algorithms, making it difficult for non-experts to independently verify results or explore new datasets. This opacity increases the risk of misinterpreting artifacts as real phenomena, as highlighted by \citet{withers2014process}. In this study, we present a comprehensive signal processing pipeline that begins with raw RO data from the Akatsuki spacecraft’s Radio Science experiment. The pipeline extracts signal intensity and frequency residuals, computes bending angles and impact parameters, and derives key atmospheric parameters; including refractive index, temperature, pressure, defocusing loss, and X-band absorptivity. The primary objective is to retrieve vertical profiles of sulfuric acid vapor, with particular focus on the 35 -50 km region. The derived sample profile is consistent with the known saturation curve of H$_2$SO$_4$, showing peak vapor abundances of a few ppm below the clouds and a rapid decline above 50 km.

\begin{figure*} [tbh]
    \centering
    \includegraphics[width=0.9\linewidth]{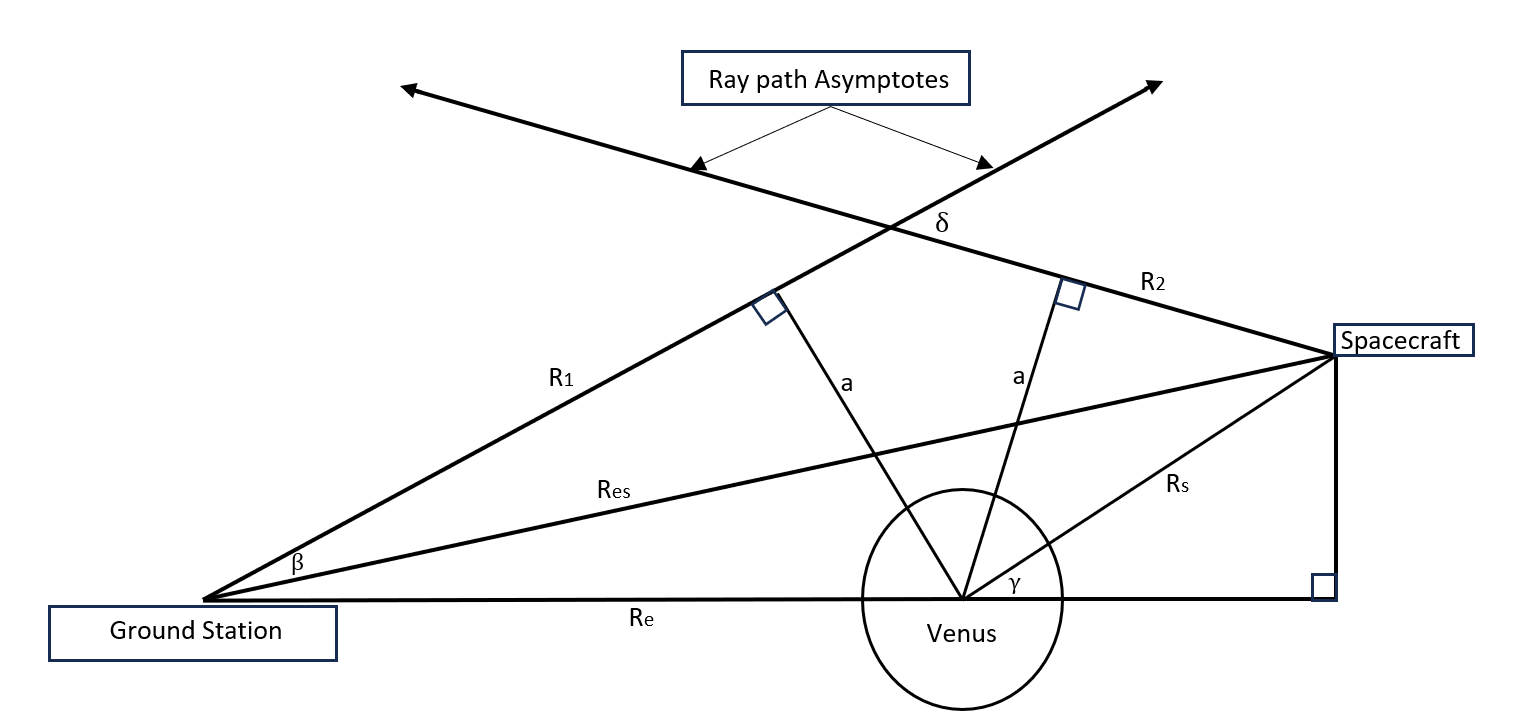}
    \caption{Schematics of the occultation geometry. Radio signal initially emitted along R$_2$ direction gets bent by an angle $\delta$  by the Venusian atmosphere and reaches the ground station on Earth. The perpendiculars to the ray-path asymptotes designated as $\textbf{a}$ the impact parameters are also shown}
    \label{fig:schematic}
\end{figure*}

\section{Methodology}
\label{sec:methodology}
The thick, visually opaque Venusian atmosphere has been difficult to probe below the cloud top regions. Radio Occultation (RO), however, offers a distinct opportunity and advantage in this regard, as we can remotely sound the deep sub-cloud altitudes, down to 35-40 km, with a significantly high vertical resolution of a few hundred meters, which other instruments cannot provide. Vertical profiles of other physical parameters, namely, temperature, pressure, neutral and plasma densities have been also been derived with near-global coverage employing this technique using Venus orbiter missions since the late 1970s like Pioneer Venus Orbiter mission (PVO), ESA's Venus Express, JAXA's Akatsuki mission, and others.

In a typical one way RO experiment like the one conducted with the Radio Science (RS) payload onboard the Akatsuki spacecraft orbiting Venus \citep{imamura2017initial}, a stable oscillator (ideally an Ultrastable Oscillator or USO) onboard the spacecraft orbiting a planetary body, emits an electromagnetic (EM) signal in the microwave domain (S, X or Ka band). The EM wave traverses the atmosphere of the target planetary body, suffers a phase change and loss in signal intensity, and reaches a receiving station on Earth. The relevant atmospheric parameters are derived by analyzing this phase change and the intensity loss \citep{imamura2017initial, tripathi2022estimation}.

\subsection{Phase Change}
\label{subsec:phase}
The phase change in the EM wave due to interactions with the intervening medium is observed as a shift in the frequency of the received signal compared to the original transmitted frequency \cite{oschlisniok2012microwave}. This frequency perturbation, along with the relevant occultation geometry, are used in an iterative loop to calculate the bending angle and the impact parameter of the radio wave during the ingress(starting)/egress(ending) phase of an occultation experiment (Figure \ref{fig:schematic}). The details of this method are described elsewhere \citep{fjeldbo1971neutral, tripathi2022estimation, tripathi2022quantification}. It has been referenced and used quite regularly by the radio science community \citep{lipa1979statistical, jenkins1994radio, patzold2007structure, oschlisniok2021sulfuric, tripathi2022estimation}. The radially varying refractive index of the Venusian atmospheric medium leads to the bending of the radio wave. In a stably striated atmosphere, the bending of a radio ray path can be approximated if the refractive index values of the atmosphere at each altitude are known from considerations of simple geometrical optics. Thus, in our case, it becomes an inversion problem where the expected input, the medium's refractive index values, are to be calculated from the already estimated \enquote{output} results, viz. the bending angles and the impact parameters. Again, following the method developed by \cite{fjeldbo1971neutral}, an onion peeling technique assuming spherical symmetry in the atmospheric system is implemented, using an Abel inversion to calculate the refractive index as a function of the ray periapsis (r$_o$), the distance of closest approach of the radio ray from the center of the planet,

\begin{equation}
   \ln(\mu(r_0)) = \frac{1}{\pi} \int_{a(r_0)}^{\infty} \frac{\delta(a) \cdot da}{\sqrt{a^2 - a(r_0)^2}}.
   \label{eq:abel_refractive}
\end{equation}
where $\mu$ is the real part of the refractive index of the medium, $\delta$ is the ray bending angle, and $a$ is the impact parameter. Now, the probed radial distance r$_0$ is calculated using Bouguer's law, $a = \mu(r_0)\cdot r_0$  \citep{born2013principles}, and the refractivity of the medium is $N = \mu - 1$.

Once we have the refractivity, it is relatively easy to calculate the neutral number density (n) of the atmosphere, using the expression, $N = \mu - 1 = K \cdot n$, where K is the mean refractive volume applicable for the Venusian atmosphere \citep{eshleman1973radio, tripathi2022estimation, tripathi2022quantification}. The K-value of $1.804 \cdot 10^{-29} m^3$ has been taken from \citet{hinson1999initial} for a 96.5\% CO$_2$ and 3.5\% N$_2$ planetary composition, which should be applicable for both Venus and Mars. As the Venusian atmosphere below 100 km is considered to be well mixed, a hydrostatic approximation can be applied to derive the temperature profile, with a pre-set composition (96.5\% CO$_2$ and 3.5\% N$_2$) and a suitable boundary condition \citep{lipa1979statistical, tellmann2009structure, imamura2017initial, tripathi2022estimation},

\begin{equation}
    T(h) = \frac{\mu_{\text{top}}}{\mu(h)} \cdot T_{\text{top}} + \frac{\overline{m}}{k_B \cdot n(h)} \int_{h}^{h_{\text{top}}} n(h') \cdot g(h')    dh'
\label{eq:placeholder}
\end{equation}
where $T(h)$ is the temperature of the atmosphere at an altitude $h$, $\overline{m}$ is the mean molecular mass of the neutral atmosphere, $k_B$ is the Boltzmann constant and $g(h')$ is the acceleration due to gravity of Venus that changes with the altitude within the integral. The boundary condition applied here is the standard value adopted by many works in the literature \citep{tellmann2009structure,imamura2017initial, tripathi2022estimation}. It assumes that at the 100 km altitude, set as the boundary of the neutral atmosphere, the temperature is 200 K. Other boundary values of 170 K and 230 K have also been considered \cite{tellmann2009structure, imamura2017initial, tripathi2022quantification}, but they do not have any impact on the temperatures below 75-80 km, and since this work is chiefly concerned with the 35-55 km regions, we can safely take the 200 K boundary value. With the already calculated temperatures, the pressure is now estimated with the help of the ideal gas equation, $P = n\cdot k_B\cdot T$.

\subsection{Intensity Loss}
The other extracted parameter from the raw data file is the signal intensity as a time series array corresponding to the ingress or the egress phases of occultation. The details of the signal processing steps to derive the said intensities are given in Section \ref{sec:signal_processing}. The stable radio transmitter onboard the spacecraft is usually turned on a few minutes ($\sim$ 10-15 mins) before the actual occultation starts to enable an accurate baseline correction to the frequency residuals \citep{imamura2017initial, tripathi2022quantification}. Initially, when the closest approach of the ray path with respect to the target planetary body (Venus) is large, such that the signal traverses the interplanetary medium only, a strong signal with small deviation (typically less than 0.5 dB in relative units) is received at the ground station. However, when the radio ray breaches the 100 km altitude, the density of the probed neutral Venusian atmospheric region becomes quite significant, and the received signal intensity starts falling drastically. It becomes quite difficult to separate the signal from the noise in the deepest parts of the atmosphere below 40 km, with the signal attenuation ending up somewhere around -40 to -45 dB in relative units. Although a theoretical refractive limit of 32 km is present for Venus \citep{tellmann2009structure}, we can only probe down to 38-40 km at best with the Akatsuki spacecraft RO experiment, which is standard for most Venus RO experiments,  including Venus Express.

There are three main factors responsible for the total signal attenuation $\phi$ -- (a) the defocusing or the refractive loss L($r_o$), (b) the spacecraft antenna mispointing loss M($r_o$), and (c) Venus atmospheric absorption loss $\tau$($r_o$) \citep{lipa1979statistical, oschlisniok2021sulfuric}.

\subsubsection{Defocusing Loss}
A radio beam traverses the target atmosphere, and there is a defocusing effect observed in the plane of refraction due to the radial dependence of refractivity. Additionally, the beam gets compressed in the plane perpendicular to the plane of refraction due to the curved limb of the atmosphere \citep{lipa1979statistical, eshleman1980comment}. In the net combined effect, the defocusing part dominates over the ray focusing, leading to a net loss in signal intensity at the receiver's end on Earth. This defocusing loss is calculated using the retrieved bending angle, the impact parameter, the probed radial distance and the applicable occultation geometry (Figure \ref{fig:schematic}) \citep{eshleman1980comment},

\begin{equation}
L(r_0)^{-1} = \frac{r}{a} \left(1 - R_2 \frac{d\delta}{da}\right)
\label{eq:defoloss}
\end{equation}
where R$_2$ is the perpendicular distance from the spacecraft to the closest impact parameter along the ray-path asymptote emanating out of the spacecraft.

\subsubsection{Mispointing Loss}
For any radio occultation experiment, and especially for the ones probing dense atmospheres like that of Venus, accurate spacecraft antenna pointing becomes essential to retrieve the signals from the deepest altitudes. Even small misalignment in the antenna boresight leads to significant degradation in the signal, and it becomes impossible to separate the signal from the noise floor \citep{tripathi2022quantification}. So for the Akatsuki RO experiments, carefully planned spacecraft slews ensured the correct pointing of the antenna boresight so that the pointing errors were minimal \citep{imamura2017initial}. For the particular experiment day considered in this work, the mispointing loss was negligible and was ignored.

\subsubsection{Absorption Loss}
\label{subsec:absorption_loss}
The other main contributor to the loss of signal intensity during an occultation experiment is the atmospheric medium of the target planetary body itself. Significant signal attenuation due to absorption effects ($\sim$15-20  dB) is observed during the X-band occultation experiments on Venus, particularly below 50 km altitude \citep{Cimino_phd, Oschlisniok_phd}. Similar trends are observed in the S-band as well, although the attenuation (in terms of $\alpha$ as expressed in Equation \ref{eq:alpha_abel}) is an order of magnitude less \citep{cimino1982composition}. While pressure-broadened CO$_2$, N$_2$, and microwave active SO$_2$ do cause some absorption, the major X-band absorber in the Venusian atmosphere is the H$_2$SO$_4$ vapor layer underneath the clouds \citep{steffes1982sulfuric, kolodner1998microwave, oschlisniok2021sulfuric, akins2023approaches}.

The total absorption loss $\tau$ is calculated by subtracting the sum of the defocusing loss L and the mispointing loss M from the total signal attenuation $\phi$.
\begin{equation}
    \tau = \phi - L - M
\label{eq:excess_attenuation}
\end{equation}
$\tau$ is a path-integrated loss quantity expressed in dB. However, in an RO experiment, since the vertical limb of the atmosphere is scanned and studied, it is more appropriate to use specific attenuation, or attenuation per unit length, expressed in dB/km. The specific attenuation $\alpha$, also known as the absorption coefficient or the absorptivity,  when integrated over the radio ray path using Bouguer's rule, results in $\tau$. An Abel inversion converts each $\tau$ value, calculated from Equation \ref{eq:excess_attenuation}, into the X-band absorptivity $\alpha$ corresponding to each altitude point \citep{jenkins1994radio, oschlisniok2012microwave},

\begin{equation}
    \alpha(r_0) = -\frac{\mu(r_0)}{\pi \cdot a(r_0)} \cdot \frac{d}{da} \left[ \int_{a(r_0)}^{\infty} \frac{\tau(a) \cdot a    da}{\sqrt{a^2 - a(r_0)^2}} \right]
\label{eq:alpha_abel}
\end{equation}

The above Equation \ref{eq:alpha_abel} provides the total X-band absorptivity where all the microwave absorbing atmospheric constituents have an impact. The contributions to $\alpha$ from the major known absorbers are calculated first. Laboratory-based studies mimicking Venus atmospheric conditions were carried out by \citet{ho1966laboratory} to study the microwave absorption by CO$_2$, N$_2$, H$_2$O, and Ar under the pressure-broadened conditions of Venus. The study was replicated by \citet{steffes2015laboratory}, and they also ended up with a similar empirical relation, given by,

\begin{multline} 
\alpha_{\text{CO}_2,\text{N}_2} = (1.15 \times 10^8) \times  \\ f^2  \left( p 1.01325^{-1} \times 10^{-5} \right)^2 \times \\  T^{-5} \left( q_{\text{CO}_2}^2 + 0.25 q_{\text{CO}_2} q_{\text{N}_2} + 0.0054 q_{\text{N}_2}^2 \right) \label{eq:alpha_CO2}
\end{multline}
where only the CO$_2$ and the N$_2$ contributions were considered.

Along with CO$_2$ and N$_2$, Sulfur dioxide vapors also play a non-negligible role in the X-band absorption. Initially in 1981, \citet{janssen1981microwave} used the Van Vleck Weisskopf theory to estimate X-band absorptivity by SO$_2$ vapors in a CO$_2$ dominated Venus-like atmosphere. Subsequent improvements were made by \citet{steffes1982sulfuric}, \citet{FAHD1992200}, and by \cite{https://doi.org/10.1029/95JE03728} using laboratory-based studies and a Ben-Reuven line shape model. \citet{oschlisniok2012microwave} used these newer studies to update the Janssen and Poynter empirical relation of X-band SO$_2$ absorptivity,

\begin{equation}
    \begin{split}
        \alpha_{\text{SO}_2} = 4.3 \times 10^6 f^2  \left( p \cdot 1.01325^{-1} \times 10^{-5} \right)^{1.28} \times \\  T^{-2.91} q_{\text{SO}_2}         
    \end{split}
\label{eq:alpha_SO2}
\end{equation}

Thus, knowing the SO$_2$ vapor concentrations in the region, the absorptivity can be calculated. The X-band absorptivities of CO$_2$, N$_2$ and SO$_2$ are then added up and they form the known component of the total absorptivity, $\alpha_{known} = \alpha_{SO_2} + \alpha_{CO_2} + \alpha_{N_2}$. Subtracting the known part from the total absorptivity, we end up with the X-band absorptivity due to H$_2$SO$_4$ vapors. \citet{kolodner1998microwave} carried out lab-based studies and provided the empirical relation to estimate the H$_2$SO$_4$ vapor abundance for a given absorptivity value applicable for Venus,


\begin{multline}
\alpha_{\text{H}_2\text{SO}_4}(3.6 \text{ cm}) = 443.570 \cdot \left( \frac{553}{T} \right)^{3.0 \pm 0.2} \times \\ \left( p \cdot 1.01325^{-1} \times 10^{-5} \right)^{1.302}  \cdot q_{\text{H}_2\text{SO}_4} \label{eq:alpha_sulfuric}
\end{multline}

While deriving the profiles, the impact of the minor absorbing constituents has been ignored. It has been shown in previous studies that the combined contribution of the minor absorbers like H$_2$O vapor, OCS, and others is less than 1\% of the total absorptivity $\alpha_{total}$ \citep{akins2023approaches}. During the occultation experiments, one implicit assumption that has gone in is the constancy of the Earth's atmospheric conditions. So even though certain constituents in the Earth's atmosphere like H$_2$O vapor will cause some X-band loss, the unchanging atmosphere should affect a steady attenuation and thus the total signal loss will be uniform throughout. The transmitted X-band signal's wavelength ($\sim$ 3.5 cm) is too large to be scattered significantly by the Venusian atmosphere, and thus any attenuation effects due to atmospheric scattering are ignored. One caveat in the estimation of the $\alpha_{known}$ component is that although we have an approximate idea about the SO$_2$ vapor abundance, large uncertainties remain. A couple of approaches, including an indirect method implemented by \citet{oschlisniok2021sulfuric}, have been used to estimate the SO$_2$ vapor abundance in the 51-54 km region of the atmosphere, whose details will be discussed in the Section \ref{sec:sulfuric_main}.

\section{Signal Processing}
\label{sec:signal_processing}
The Akatsuki spacecraft, launched in December 2010 and placed in the Venusian orbit in 2015 after an initial failure, conducted Radio Occultation experiments from 2016 to 2024. The payload, equipped with an Ultrastable Oscillator (USO), emitted in the X-band ($\sim$8.41 GHz) with a stability of $\sim10^{-12}$ over $1-1000$s of integration time, which corresponds to an uncertainty of $\sim$$0.01$Hz in the transmitted signal \citep{imamura2017initial, tripathi2022quantification}. The radio science receivers (RSR) of three independent ground-based deep space network (DSN) stations were used for tracking the spacecraft on different days, so that for some of the experiment days, simultaneous observations from two of the three stations are available. The three DSN tracking stations were at (i) UDSC of JAXA in Japan, (ii) IDSN of ISRO in India, and (iii) DLR Weilheim Aerospace station in Germany. For this study, we have considered one of the RO experiment days recorded at IDSN, India. The RO experiment was conducted in the Open Loop (OL) mode, which provides a clear advantage over the Closed Loop (CL) mode when probing highly dense planetary atmospheres as it enables the use of advanced digital signal processing (DSP) techniques to extract data from deep altitudes of the target atmosphere ($<50$km for Venus) \citep{imamura2017initial, tripathi2022estimation}.

\begin{figure*}[tbh]
\centering
\begin{tabular}{cc}
\includegraphics[scale=0.25]{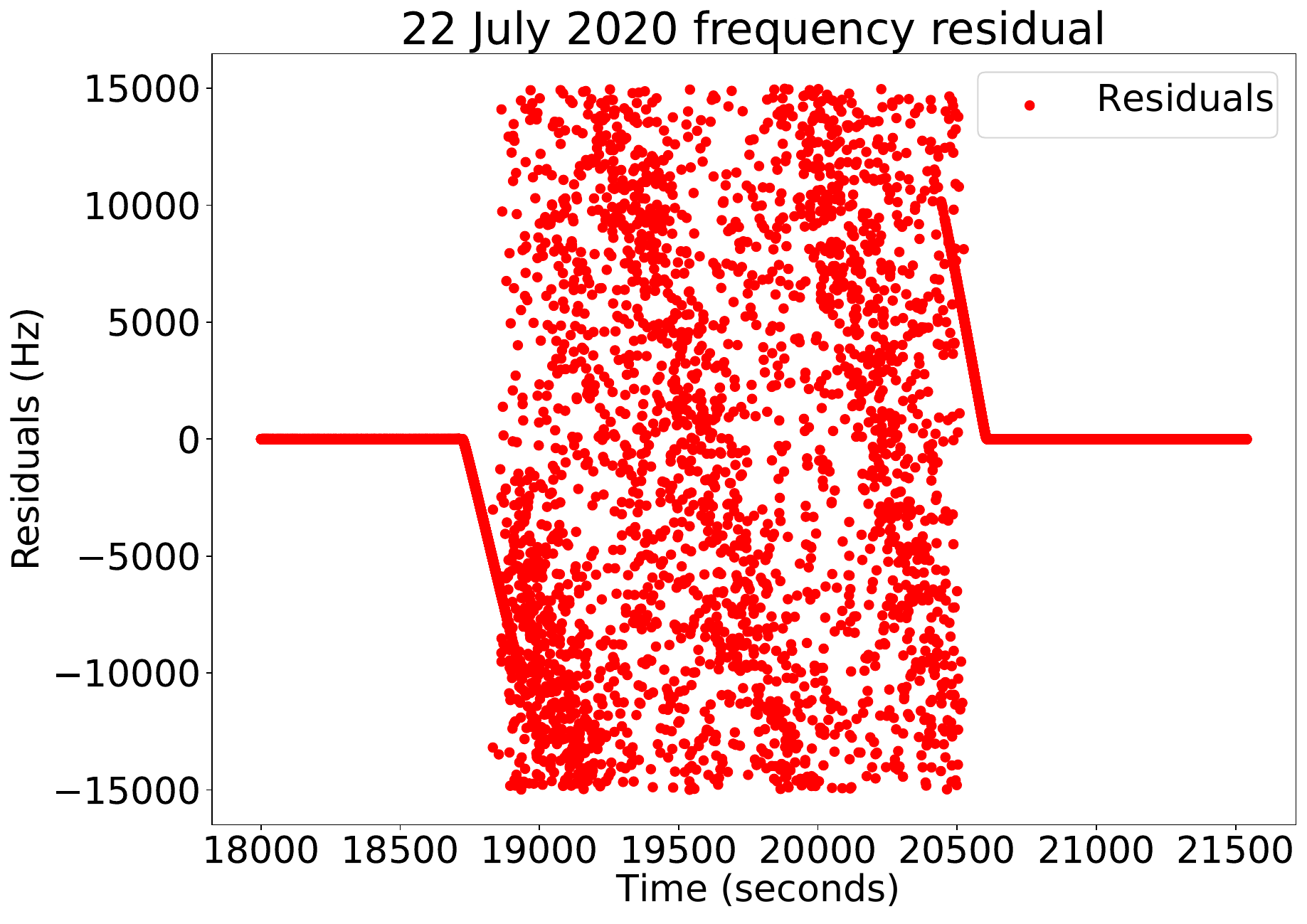} &
\includegraphics[scale=0.25]{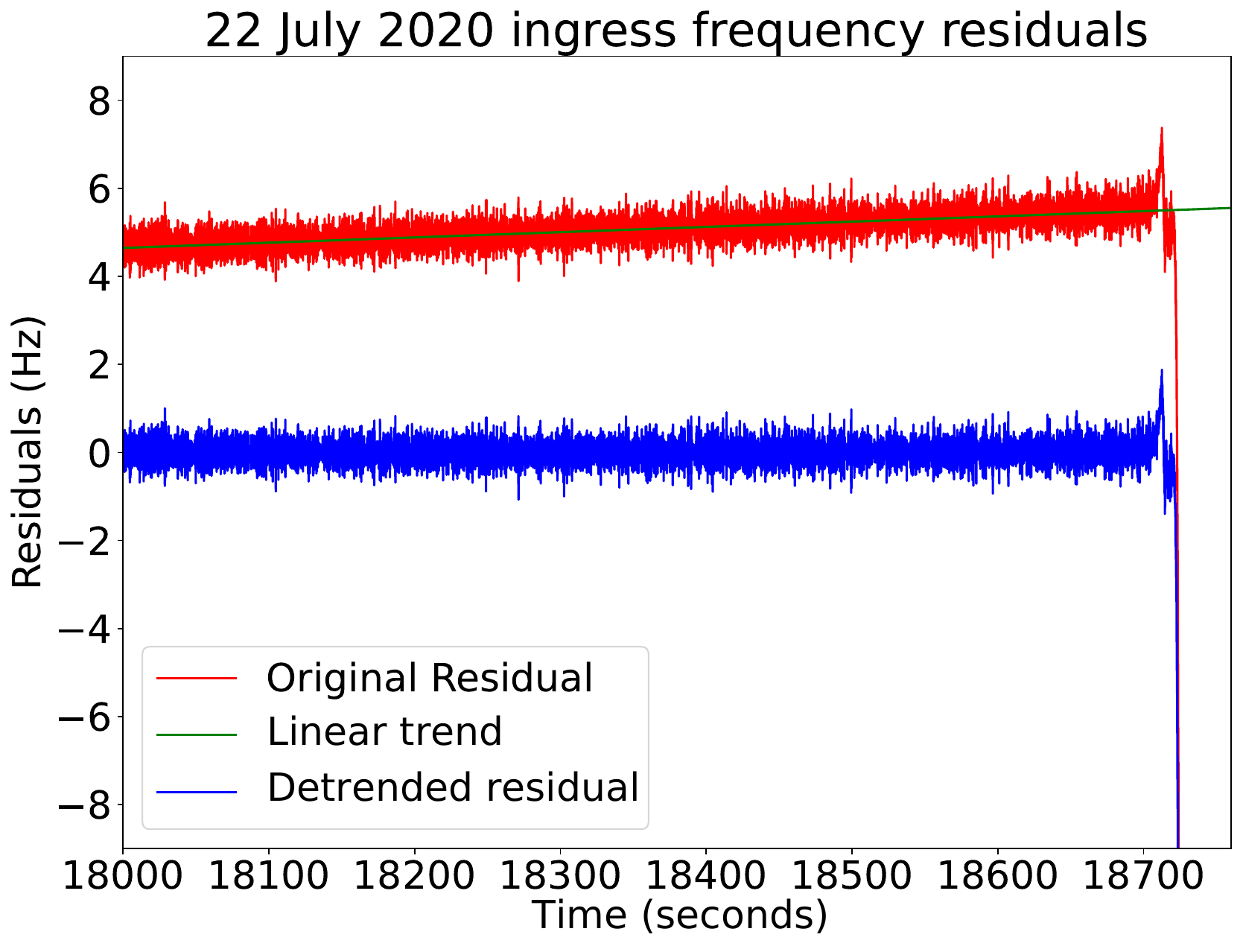} \\
\includegraphics[scale=0.25]{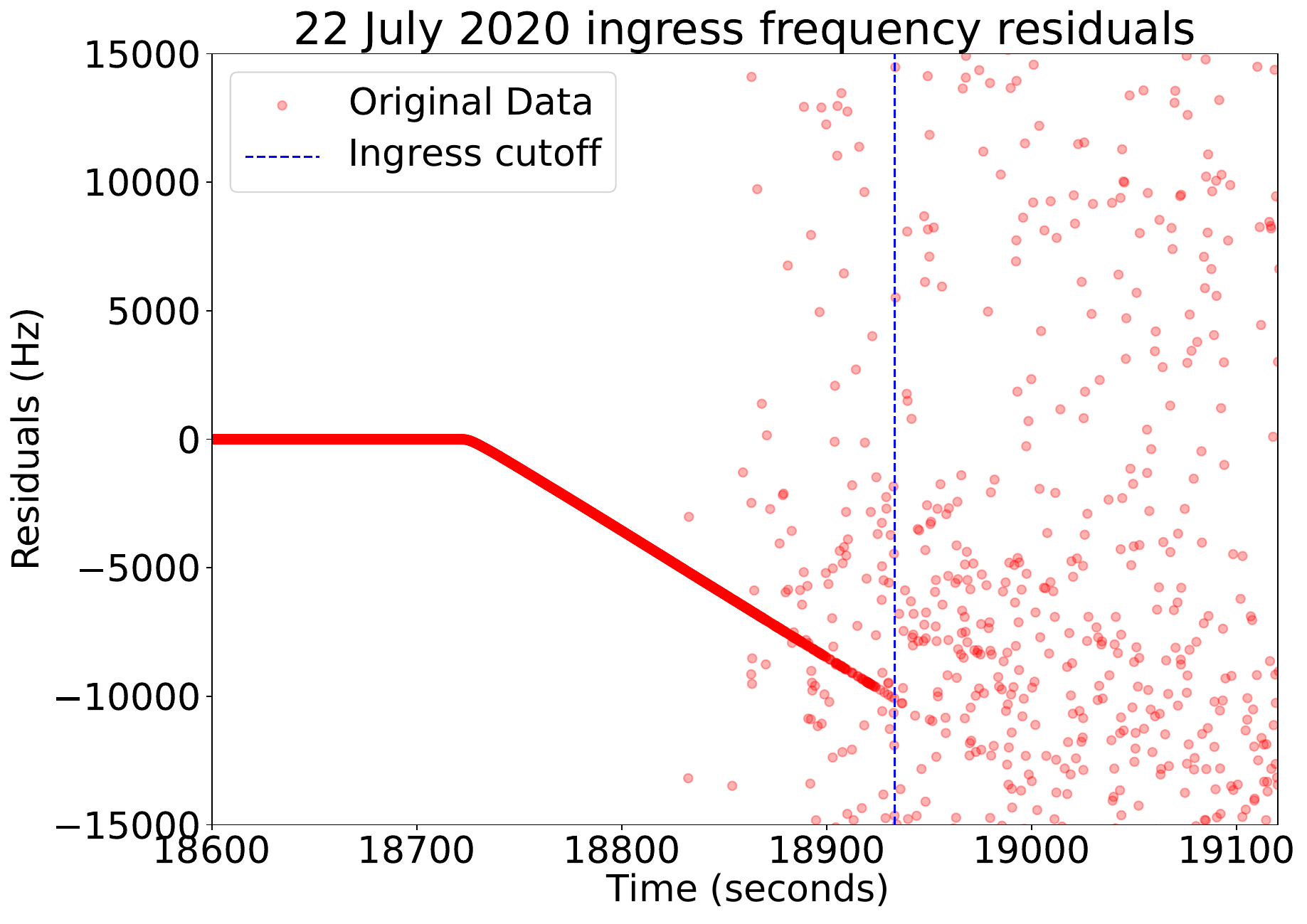} &
\includegraphics[scale=0.25]{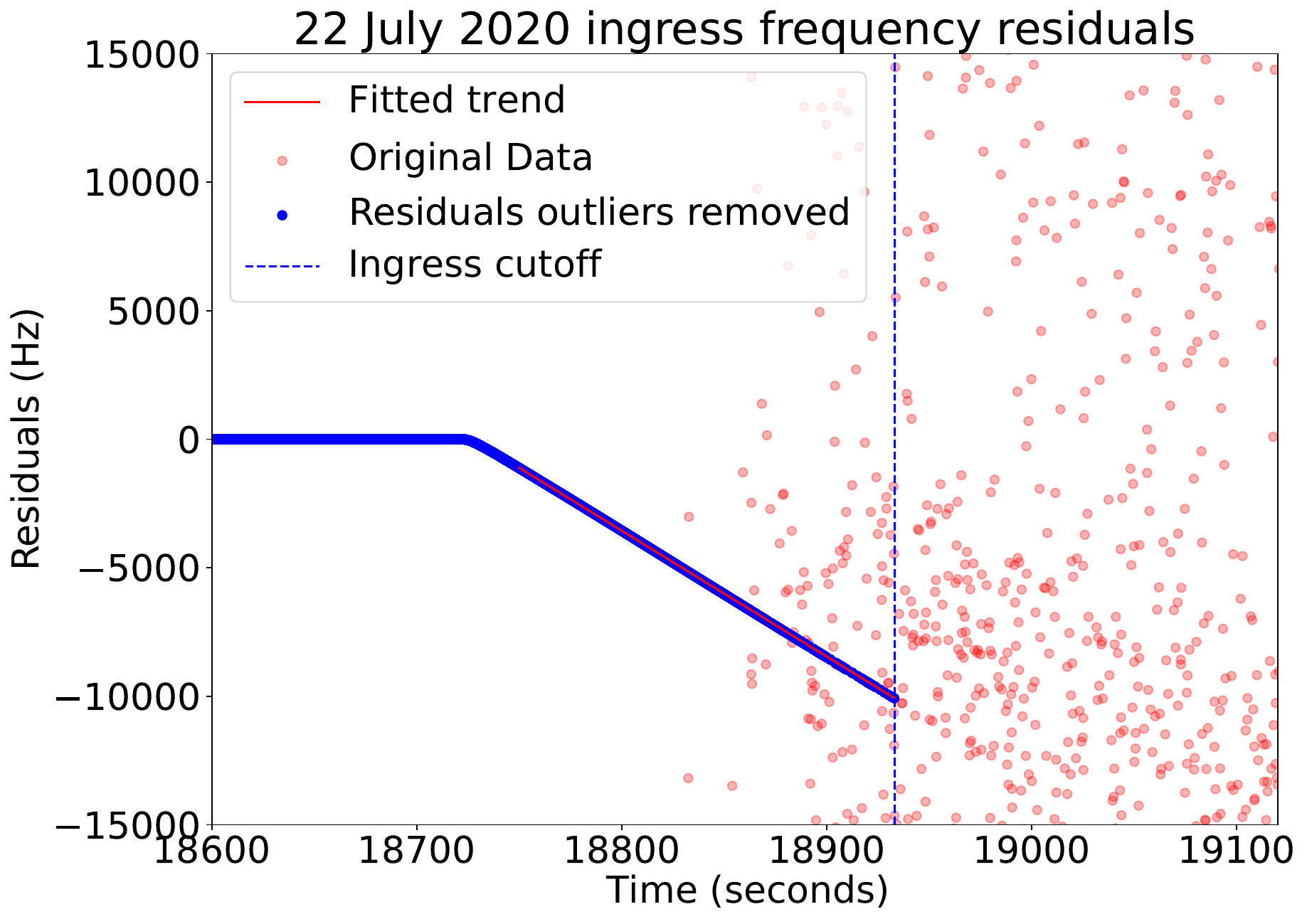}
\end{tabular}
\caption{Frequency Residuals for 22 July 2020 ingress. The first figure (top left) shows the combined residuals of the ingress and egress phases of the occultation experiment. The second plot (top right) shows the linear trend in the residuals (green line) often observed in planetary occultation experiments primarily due to prevailing background conditions, and how a linear baseline correction is done to remove its effect. Next, (bottom left) it is seen that even though large fluctuations in the residuals exist, the predominant linear (or non-linear) trend in the residuals in the deep atmosphere is still maintained even through the fluctuations. A linear trendline (red line), as shown in the bottom right plot, in the non fluctuating region of the lower atmosphere (between $18750s$ and $18840s$) is extrapolated forward in time till the ingress cutoff (blue dashed line) is reached, and a threshold of around 40 Hz is used to pick out the \enquote{correct residuals} from the data. }
\label{fig:resi}
\end{figure*}

The OL mode of RO experiments with the Akatsuki spacecraft and the IDSN ground station involves the reception of the X-band signal by the RSR at IDSN and its downconversion with the help of a local oscillator (LO) to an intermediate frequency (IF) of 70 MHz using a superheterodyne technique. This makes it easier and less cost-intensive to further process and digitize the signal instead of directly digitizing the GHz wave. The downconverted signal is then sent to a signal processing center (SPC), where it is further downconverted using another LO to baseband. The expected shift in the carrier signal frequency induced by the target atmosphere is $<100$kHz during a Venusian occultation experiment, as is evident from Figure \ref{fig:resi} bottom right panel, and hence, a sampling rate of 200 kHz employed at IDSN to digitize the signal is acceptable. Finally, the raw data is stored in the complex form of in-phase (I) and quadrature (Q) components as a time series array \citep{tripathi2022estimation}. The data is recorded at IDSN, India, in the raw data exchange format (RDEF), which has been developed by NASA. Details about the RDEF and the method of extraction of data from the RDEF file can be found in the CCSDS blue book (https://public.ccsds.org/Pubs/506x1b1.pdf). Every second's data is sequentially stored in the raw file, each containing a header section and a data section. The header contains relevant information such as the sampling rate, RF to IF downconversion frequency, the datetime value (in UTC) of each particular moment the signal is received, etc., and the data section is an array of 200,000 sets of I and Q values.

One second's data is extracted at a time, and upon processing the entire 200,000 data points together that are present in the second, a broadening of the power spectrum is observed. This undesirable broadening is primarily brought about by the change in the line of sight (LOS) velocity between the transmitter (Akatsuki spacecraft) and the receiver (IDSN, on Earth) and it severely hinders the accurate estimation of the peak of the power spectrum, which is crucial for the retrieval of atmospheric parameters using the Radio Occultation technique \citep{tripathi2022estimation}. This problem has been addressed in detail by \citet{tripathi2022estimation}, and a similar approach has been adopted to solve the Doppler broadening issue in this work. When the data is binned at a higher time resolution ($\sim0.082s$), taking $2^{14}$ (16384) points in each bin, the spectral broadening effect goes away, and a sharp peak in the power spectrum plot is seen. The sliced data bin is then zero-padded (padding of $2^{19}$) to improve the frequency resolution, and we end up with $\sim12$ data packets for every 1 second of data. Each slice is extracted and processed independently using standard Digital Signal Processing (DSP) techniques. The reason for taking the optimized slice size of $2^{14}$ and the zero padding of $2^{19}$ is discussed in \citet{tripathi2022estimation}. It should be noted that if a different sampling rate is used during an RO experiment, the same slicing and zero padding scheme is unlikely to provide useful results and might need to be updated. This aspect has not been explored in this work.

\subsection{Estimation of Total Signal Attenuation}
Each slice of data, containing $2^{14}$ sets of I-Q points, is extracted from the raw file, and the DC bias is removed following \citep{tripathi2022estimation}. DC bias (or DC offset) is a constant voltage shift in the radio signal that mainly occurs from local oscillator leakage in mixers, power supply noise, etc. and is calculated by subtracting the mean of the raw signal from the I and Q data for each second. It is subsequently converted into a complex signal array given by $I+jQ$. After zero padding the signal, the Fast Fourier Transform (FFT) technique is applied to each data slice. FFT is one of the most popular signal processing algorithms that is used to find which frequencies in an EM signal play the dominant role. The FFT is then converted to the power spectral density (PSD), where $PSD = (1/N)\cdot(abs(FFT))^2$, N being the sample size, and abs(FFT) is the modulus of the evaluated FFT. PSD tells us how the power of the signal is distributed across different frequencies. The goal is to find the intensity and the Doppler-shifted frequency of the incoming signal. A narrow band pass filter is used to extract 512 points symmetrically around the peak of the PSD. This window of 512 points is used for further analysis for each slice of data.

Next, the noise floor of the PSD is estimated. Initially, the radio ray passes through the interplanetary medium (IPM) only and a strong signal is received at the ground station with a clearly discernible peak. However, as the dense atmosphere is probed, the received signal intensity drops exponentially and it becomes increasingly difficult to identify the peak from the noise. To get a more accurate estimate of the PSD, the denoising scheme first given by Hildebrand and Sekhon in 1974 \citep{hildebrand1974objective} is applied to calculate the constant noise floor of the PSD at each instant. The ratio R = ($mean^2/Variance$) of the sorted PSD is used to set the noise floor criteria, as an iteration is implemented to cumulatively estimate the mean and the variance of the PSD starting from the lowest values of the sorted PSD, until the condition $R\geq1$ is reached, where the actual signal emerges from the noise. The averaged sum of the PSD up to that iteration (where $R<1$) is taken as the noise power $P_{noise}$. The windowed PSD is then denoised using $P_{noise}$ thus calculated above. A Gaussian fit is applied to the denoised PSD and its statistical moments are calculated using the formalism of \citep{woodman1985spectral}. The zeroth moment is the total power, the first moment gives the Doppler shift of the received spectra, and the second moment is the spectral width of the PSD. In this work, only the zeroth and the first moments are used for the retrieval of the atmospheric parameters. 

To get a better estimate of the signal-to-noise ratio, especially at the lowest altitudes, the integration time is increased to 1 second, as suggested in \citep{imamura2017initial}, and the total power of the received signal is calculated.

\begin{figure*}
    \centering
    \includegraphics[scale=0.35]{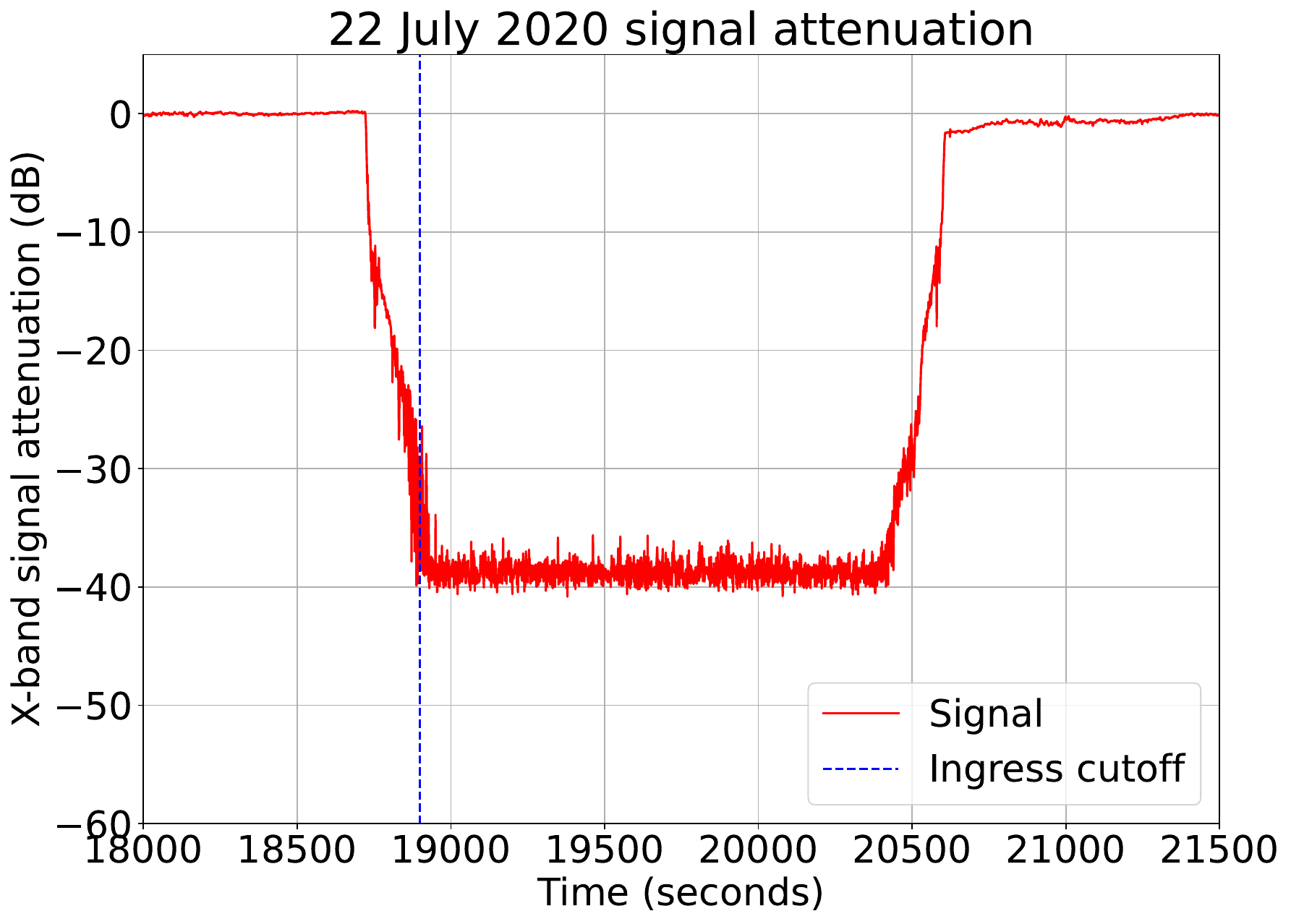}
    \caption{Signal Attenuation time series data for 22 July 2020. The signal (red curve) is steady and strong in the initial stages of the experiment and starts falling drastically when the probed altitude goes below 100 km, at 18720s. The ingress cut-off is marked by the blue dashed line, after which only the noise floor is seen in the power spectrum until 20442s, when the signal again emerges from the noise at the egress phase of the occultation experiment.}
    \label{fig:attenuation}
\end{figure*}

Figure \ref{fig:attenuation} shows the attenuation profile retrieved by processing the raw data file for the 22 July 2020 Akatsuki RO experiment of Venus. The processing technique implemented is as described above. In the first phase of the experiment, while probing the IPM and the ionospheric regions of Venus (above 100 km), the signal strength is observed to be quite high, with negligible degradation. However, from $18720s$, when the ray path starts probing below the 100 km altitude of Venus, the signal intensity drops significantly. Near $18933s$, the signal bottoms out and reaches a saturation value, which corresponds to the noise floor level. This is the inaccessible part of the occultation experiment, where the high signal attenuation makes any observations below a certain altitude practically impossible. After that, the spacecraft gets completely occulted, and no observation is possible. The saturated attenuation value is approximated of the order of $-35$ dB which is in the range as obtained in earlier works \citep{imamura2017initial}.

\begin{figure}[tbh]
\centering
\begin{tabular}{c}
\includegraphics[scale=0.25]{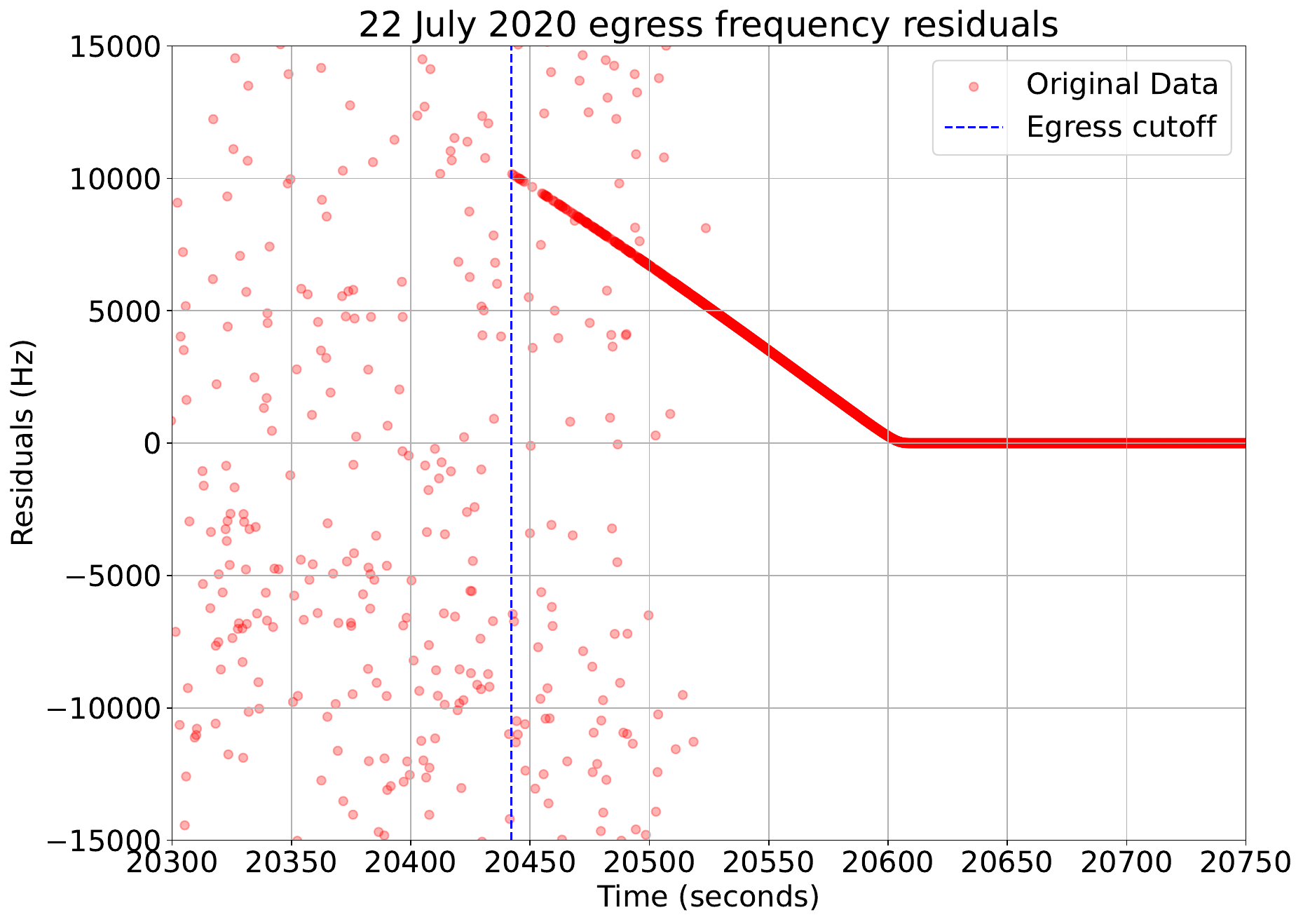} \\
\includegraphics[scale=0.25]{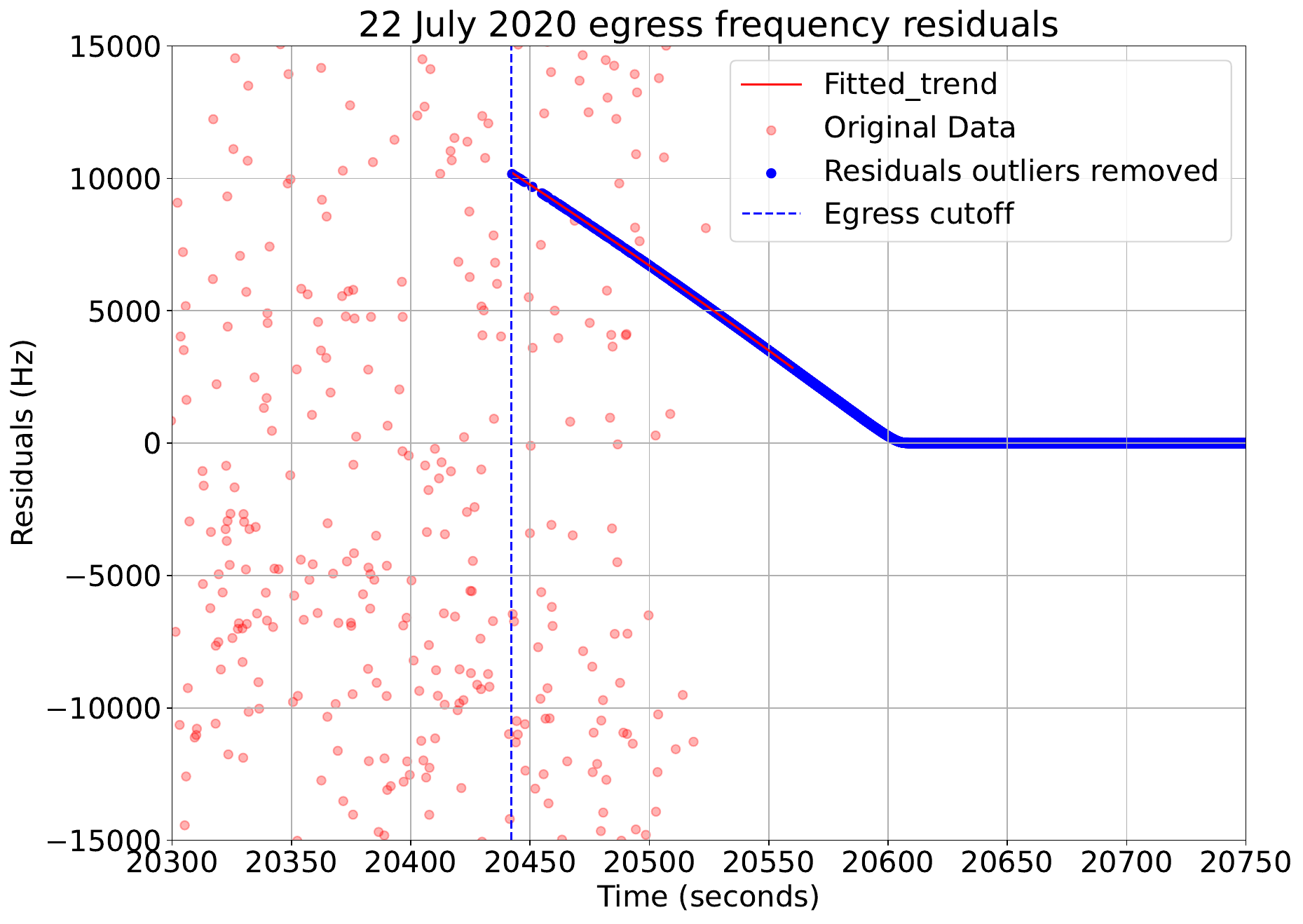}
\end{tabular}
\caption{Frequency Residuals for 22 July 2020 Egress. The bottom panel shows the non linear trend in the residuals, that has been captured by using a second order polynomial fit in the non-fluctuating regions of the egress experiment when the ray is still probing the neutral Venusian atmosphere.}
\label{fig:resi-egress}
\end{figure}

\subsection{Calibration and Processing of Frequency Residuals}
\label{subsec:calibration_resi} 
The first statistical moment of the denoised PSD gives the Doppler-shifted frequency that is received by the RSR at IDSN. This relatively large frequency shift ($\sim$350 kHz observed for the 22 July 2020 experiment) is primarily due to the relative motions of the transmitter (Akatsuki spacecraft) and the receiver (IDSN on Earth) in the inertial J2000 reference frame. The other contributing factor to the Doppler shift is, of course, the target body's atmospheric medium itself, which is to be studied here. The frequency shift due to the relative motion of the planetary bodies and spacecraft can be theoretically estimated, the details of which are given in \citet{tripathi2022estimation}. Essentially, higher order (up to $c^{-2}$) effects, namely (a) special relativistic effects due to the changing velocities of the transmitter and receiver, and (b) general relativistic effects (viz. the gravitational redshift) calculated using the gravitational potentials of Venus and Earth as encountered by the EM wave, are considered to help estimate the shift in the frequency of the transmitted photon by the stable spacecraft oscillator in the J2000 frame. The expression comes out to be as in Equation \ref{freq} after \citet{krisher1993parametrized},

\begin{multline}
\label{freq}
f_r = f_e \Big[ 1 + \hat{\mathbf{n}} \cdot (\mathbf{v}_e - \mathbf{v}_r)/c - (v_e^2 - v_r^2)/2c^2 - \\ (\hat{\mathbf{n}} \cdot \mathbf{v}_e)(\hat{\mathbf{n}} \cdot \mathbf{v}_r)/c^2 + (\hat{\mathbf{n}} \cdot \mathbf{v}_e)^2/c^2 - (U_e - U_r)/c^2 \Big]
\end{multline}

where the suffixes $e$ and $r$ stand for the emitter and receiver, respectively. $f$ is the frequency of the wave, $v$ is the velocity in the  J2000 frame, $\hat{n}$ is the unit vector pointing along the transmitter-receiver direction, and U$_e$ and U$_r$ are the gravitational potential at the transmitter's and the receiver's ends respectively. 

The positions and velocities of the spacecraft and the planetary bodies in the J2000 reference frame for the relevant time of the occultation experiments are determined using the SPICE toolkit developed and maintained by NASA's Navigation and Ancillary Information Facility (NAIF) \cite{acton1996ancillary}. The mission-specific kernels used for the accurate determination of spacecraft and the receiving station ephemerides have been provided by JAXA and IDSN teams, respectively.

The theoretically calculated Doppler shift is subtracted from the experimental Doppler value estimated after processing the raw data file. Experimental Doppler = first statistical moment of PSD + downconversion frequency - transmitted frequency.
This subtracted element chiefly comprises the frequency perturbations due to the target planetary atmosphere and ionosphere, and is termed the frequency residual ($\Delta f$),

\begin{equation}
    \Delta f = Experimental\_Doppler - Theoretical\_Doppler
\end{equation}

The frequency residuals for 22 July 2020 are plotted as a time series data and shown in Figure \ref{fig:resi} (top left panel). Initially, when the X-band signal traverses the IPM and the ionosphere of Venus, the fluctuations in the residual values are quite small. However, a linear trend is observed in the residuals (Figure \ref{fig:resi}, top right panel). This trend is most likely caused by background sources, like the terrestrial atmosphere/ionosphere, the IPM along the ray-path, inaccuracy in the precise determination of the spacecraft ephemeris, or others \cite{withers2010prediction, tripathi2022quantification}. These sources causing the linear drift in the residuals are taken to be unchanging during the whole occultation experiment, and a linear baseline correction is used to remove their impact, by taking the first 10-15 minutes of the occultation data as the reference before the target atmospheric medium is probed, as discussed in \cite{tripathi2022estimation}.

Baseline corrected residuals for 22 July 2020 are plotted in Figure \ref{fig:resi} (top right panel). The data from the first few minutes at the start of the experiment are not considered since the signal is often seen to fluctuate at the initial stages, and in this particular experiment, data from the 18000th second of the day is taken. At first, the small residual values are centered around zero. However, from the $\sim$ 18720 second value, after the ray periapsis breaches the 100 km altitude, a clear, sharp linear downward trend in the residuals is seen. As the ingress part of the occultation progresses, below 50 km near 18830s, large fluctuations are observed. It would initially appear that the signal is irretrievable beyond this time. However, on zooming in, the initial downward trend is observed to continue even through the large fluctuations until 18933s, after which the signal gets completely buried in noise, and no information can be extracted during the ingress from that point onward (Figure \ref{fig:resi}, bottom left panel).

To get a precise estimate of the impact of the deep Venusian atmosphere on the phase of the radio signal, it is imperative to remove the outliers or the noise from the residual values. While analyzing the ingress, a linear fit is applied to get the trend in the non fluctuating residuals corresponding to the regions below 100 km (18720s $<$ time $\leq$ 18830s) (Figure \ref{fig:resi}, bottom right panel), and then this trend is extrapolated into the fluctuating regions up until 18933s, after which the signal gets buried in the noise, as seen in Figure \ref{fig:attenuation}. Now a threshold of $\pm$40 Hz around this linear trendline (red) is used to pick out the correct residuals from the noisy data. Any subsequent data gaps that may occur due to the removal of the outliers are replaced by interpolated values. Next, the residuals are smoothed using an 11-point Savitsky-Golay linear smoothing process to remove small-scale random fluctuations. The outlier-removed smoothed frequency residuals are used for further analysis. It should however be noted that the linear downward trend in the residuals as observed for 22 July 2020 ingress is not ubiquitous. For other occultation days, non linear trends may also be seen, the best fit for which is obtained by applying a polynomial fit yielding the best estimate.

A similar pattern is observed during the egress as well (Figure \ref{fig:resi-egress}). Slowly emerging from the geometrical shadow of Venus, from the heavy noise region, a clear trend is seen from the $\sim$ 20442 second. And beyond 20524 second, the non-fluctuating residual data is observed, where the raypath probed radial distance rises above 50 km altitude. In this case, a second order polynomial fit was used to capture the trend, which was not linear. Further details on the choice of the polynomial order is given in the error section (Section \ref{sec: Error}).

\subsection{Averaging across the First Fresnel Zone}
In this work, principles of geometrical optics (GO) are adopted to study the Venusian atmosphere. In GO, wave theory approximates to a ray theory, as the wavelength of the electromagnetic wave is considered negligible. The first Fresnel zone (FFZ) dictates the vertical resolution of the derived parameters as the frequency residuals are averaged within the diameter of this FFZ, which is given by:

\begin{equation}
    d = 2\sqrt{\lambda \cdot D \cdot L} 
\end{equation}
where $d$ is the diameter of FFZ, $D$ is the distance between the transmitter and the occultation point, and $L$ is the defocusing loss, given in Equation \ref{eq:defoloss}.

In the FFZ, the signals in the same phase arrive at the receiver's end, interfere constructively, and contribute the maximum amount to the received intensity. Although this averaging leads to a poorer vertical resolution ($\sim$ a few hundred meters) of the atmospheric profiles, it increases the reliability and the robustness of the results.  These FFZ averaged frequency residuals are finally used for the derivation of the atmospheric parameters.

The bending angles and the impact parameters of the radio ray-path are re-estimated using the FFZ averaged residuals, along with the relevant occultation geometry plotted using the SPICE module. The refractive index of the medium, the defocusing loss of the signal, and the temperature and pressure profiles are derived using the method described in Section \ref{sec:methodology}. Small-scale sub-FFZ level fluctuations, which affect the accurate determination of the X-band absorptivity profiles, are nullified by a 1 km averaging of all the relevant parameters. This final averaged dataset is used to derive the H$_2$SO$_4$ vapor profiles in the 35-50 km altitudes of Venus.

\begin{figure}[tbh]
\centering
\begin{tabular}{c}
\includegraphics[scale=0.25]{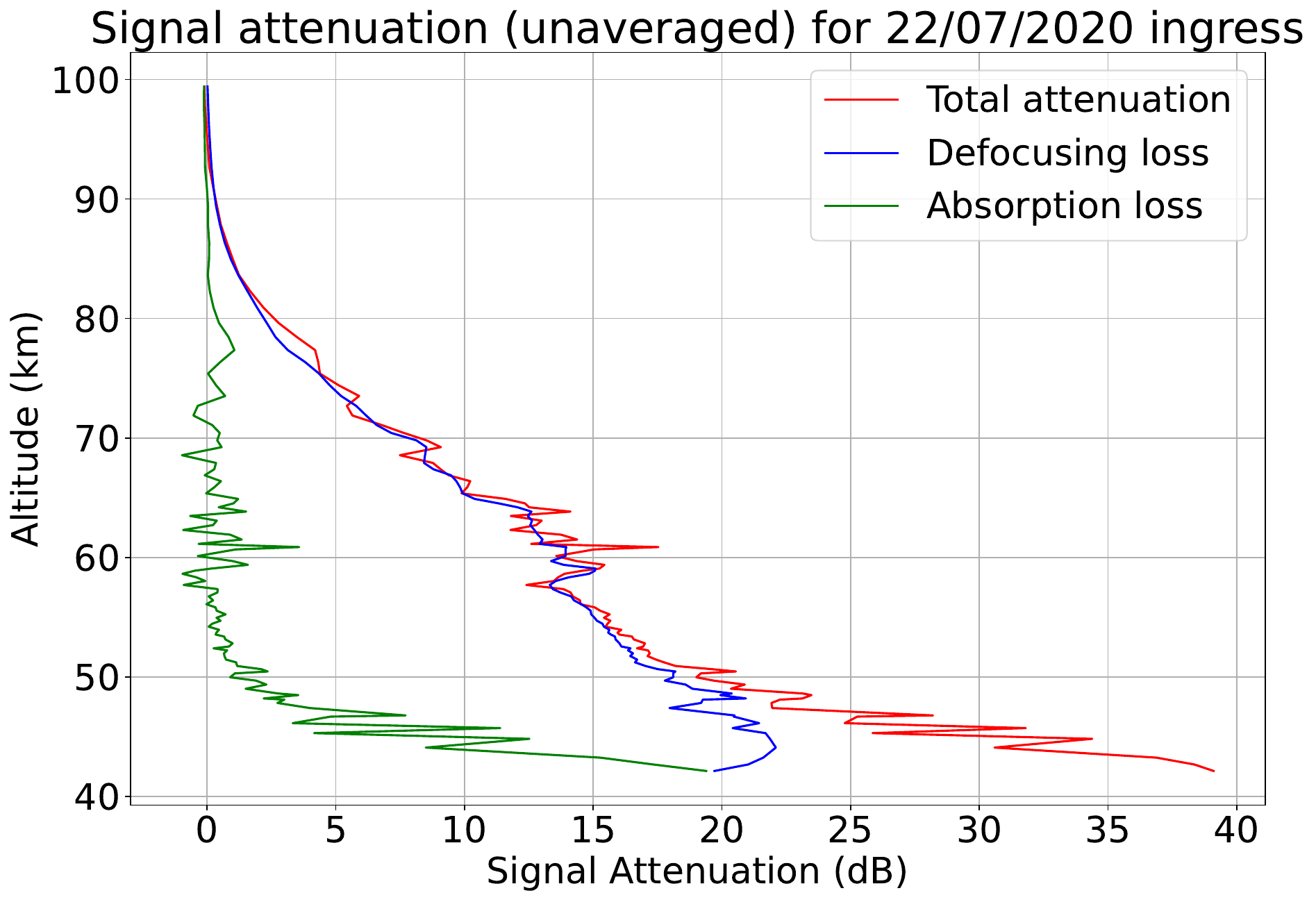} \\
\includegraphics[scale=0.25]{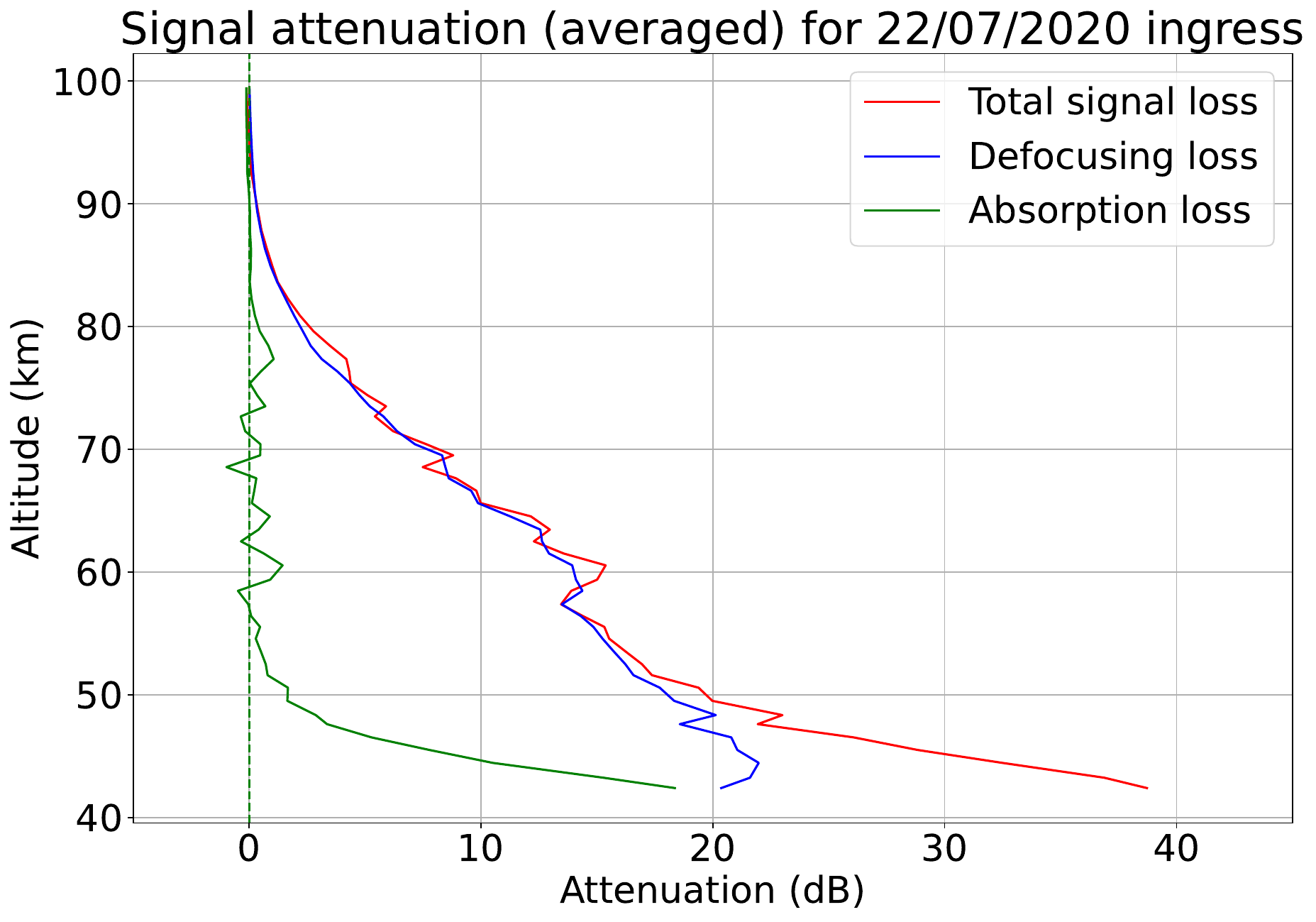}
\end{tabular}
\caption{Signal Attenuation Altitude Profile Unaveraged (top panel) and 1 km Averaged (bottom panel) for 22 July 2020 ingress. The red curve is the total signal attenuation ($\phi$), the blue is the defocusing loss (L) and the total absorption loss $\tau$ = $\phi$ - L, is represented by the green curve. $\tau$ is clearly seen to increase exponentially below 50 km altitudes, primarily due to the Sulfuric acid vapors present below the clouds.} 
\label{fig:22Jul-atten}
\end{figure}

\begin{figure*}
    \centering
    \includegraphics[scale=0.35]{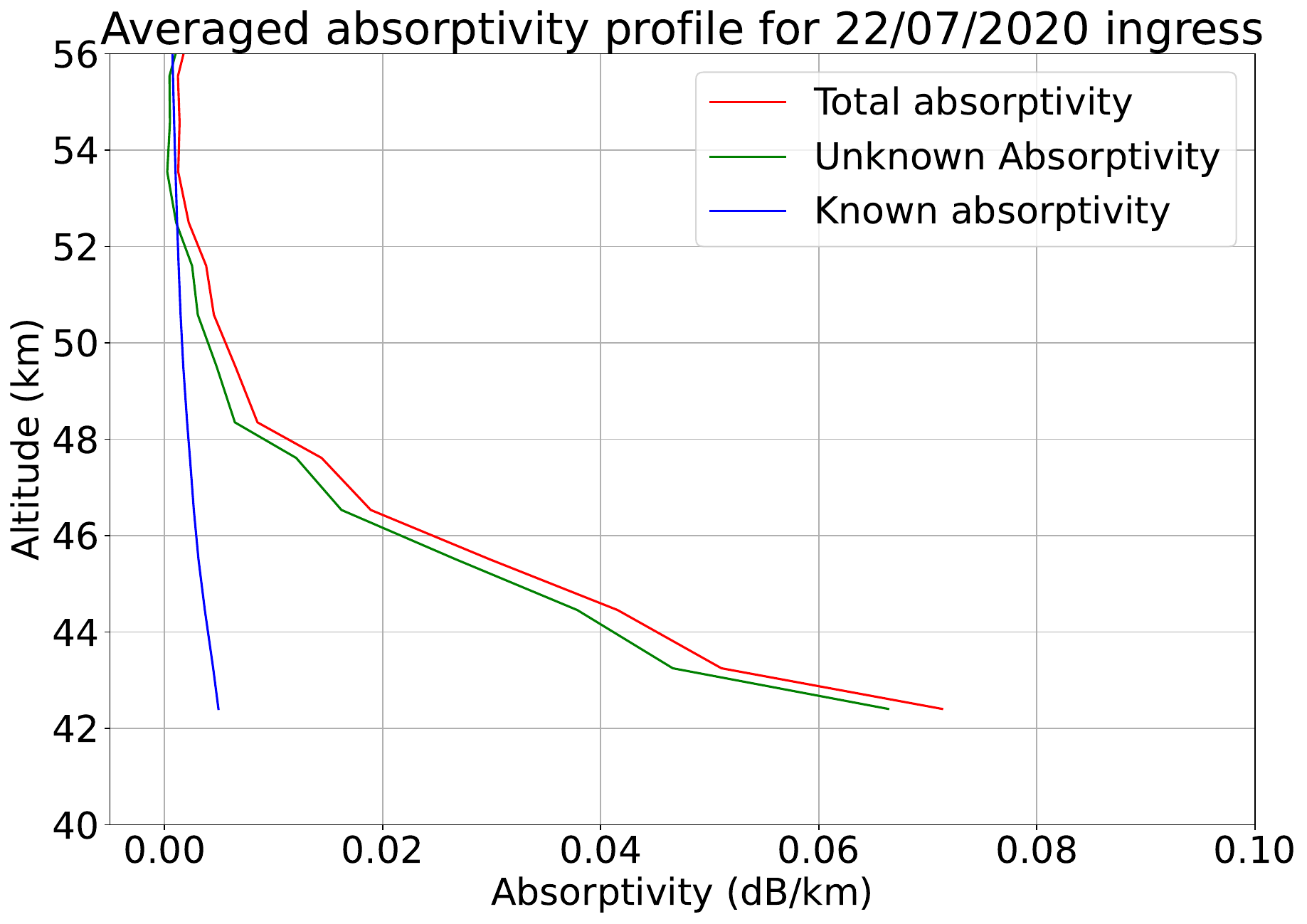}
    \caption{X-band Absorptivity ($\alpha$) Profile. The red curve is the total $\alpha$, the blue curve is the absorptivity due to the known components ($CO_2$ and $N_2$), which are considered to be well mixed in the atmosphere. Subtracting the known component from the total $\alpha$ gives X-band absorptivity due to sulfuric acid vapors. The contribution in the absorption due to the $SO_2$ vapors is calculated in Section \ref{sec:sulfuric_main}.}
    \label{fig:absorptivity}
\end{figure*}

\section{\texorpdfstring{H$_2$SO$_4$}{H2SO4} \ vapor Profiling\ and\ Evaluation\ of\ \texorpdfstring{SO$_2$}{SO2}\ vmr}
\label{sec:sulfuric_main}
The X-band attenuation altitudinal profiles (unaveraged and 1.0 km averaged) are shown in Figures \ref{fig:22Jul-atten} (top and bottom panels). The red curves represent the total signal loss, the blue ones are the estimated defocusing loss, and the green curves are the absorption loss due to the Venusian atmosphere, calculated by subtracting the defocusing loss L from the total attenuation $\phi$.

These profiles, similar to what was discussed in \citet{oschlisniok2012microwave} with VEX RO data,  can be roughly split into two regions  - above and below 55 km. It is quite clearly seen that above 55 km, the $\phi$ value is almost equal to the defocusing loss. The small fluctuations near the 60-70 km altitudes in $\phi$ are primarily attributed to upward propagating gravity waves causing the radio scintillations \citep{leroy1995convective}. 1 km averaging of the profiles (Figure \ref{fig:22Jul-atten}, lower panel) takes care of these fluctuations for the most part, and we get a clearer picture of the absorption loss, which increases exponentially below 50 km altitudes. This total absorption loss is then converted into X-band absorptivity values through the Abel inversion expressed in Equation \ref{eq:alpha_abel}. The absorptivity due to the known components is subtracted from the total $\alpha$, thereby giving us the X-band absorptivity of H$_2$SO$_4$ vapors. The absorptivity profile is shown in Figure \ref{fig:absorptivity}. Finally, the H$_2$SO$_4$ vapor abundance is calculated from the $\alpha$ values using Equation \ref{eq:alpha_sulfuric}.

The saturation vapor abundance curve is also plotted, along with the experimentally derived real abundance values, using the Clausius-Clapeyron equation appropriately tuned for Venus conditions, given in \citet{ayers1980vapor}, and then improved upon initially by \citet{kulmala1990binary} and subsequently by \citet{krasnopolsky1994h2o}. It is generally accepted that the H$_2$SO$_4$ vapor abundance follows the saturation abundance curve, and thus should become vanishingly small above 50 km regions, where the base of the clouds is typically present \citep{oschlisniok2021sulfuric}. This trend is observed in the derived profile as well (Figures \ref{fig:vega} and \ref{fig:sulfuric}). The vapor profile derived using the 22 July 2020 ingress data, observed over the latitude of -43\textdegree follows the saturation curve till $\sim$48 km, the expected boundary of the clouds. The vapor abundance is seen to increase below this height also, reaching a value exceeding 15 ppm at $\sim$42.5 km altitude. The increase in the vapor abundance below 45 km in the mid latitude regions follows the trend discussed in \citep{oschlisniok2021sulfuric}, although the peak value does show deviation from the averaged Venus Express profile corresponding to the said latitude. This deviation in abundance can be explained through the potentially large dispersion in the profiles, as observed for the equatorial regions in the case of Venus Express \citep{oschlisniok2021sulfuric}.

Sulfur dioxide (SO$_2$) is the dominant sulfur-bearing species in the Venusian atmosphere \citep{oyama1979venus}, with concentrations reaching tens and a few hundred ppms below the cloud deck \citep{oschlisniok2021sulfuric}. However, direct detection in this region is impeded by high aerosol opacity. Constraints on SO$_2$ abundance below the clouds have been primarily derived from (a) \textit{in-situ} gas chromatographic data from Pioneer Venus, (b) nephelometers onboard the ISAV 1 and 2 descent probes of the VEGA missions and (c) ground based observations of the microwave opacity and the NIR spectral windows on the night side of Venus \citep{oyama1979venus, bertaux1996vega,janssen1981constraints, bezard1993abundance, pollack1993near, marcq2008latitudinal, arney2014spatially}. \\
While contemporaneous SO$_2$ profiles are unavailable for the radio occultation (RO) observations, the species contributes appreciably to X-band signal attenuation and must be incorporated into the retrieval framework. To address this, we adopt two different approaches based on (i) in-situ data from VEGA missions and (ii) an indirect technique proposed by \citet{oschlisniok2021sulfuric}. The two methods show different SO$_2$ vapor abundances and thus the resulting sulfuric acid vapor profiles also vary.

\subsection{\texorpdfstring{SO$_2$}{SO2}  correction based on in-situ data from VEGA 1 and 2 missions}
The only available in-situ vertical profiles of SO$_2$ vapors in the lower atmosphere of Venus (below the clouds) are the ones derived from the ISAV 1 and 2 payload data of the Soviet-led VEGA missions which were launched in 1984 and entered the Venusian atmosphere on June 1985 \citep{bertaux1996vega}. The descent probes entered the atmosphere at near equatorial regions ($\sim 7-8\degree$ N and S) and the two resulting profiles only covered the low latitudes.

\begin{figure*}[tbh]
\centering
\begin{tabular}{cc}
\includegraphics[scale=0.25]{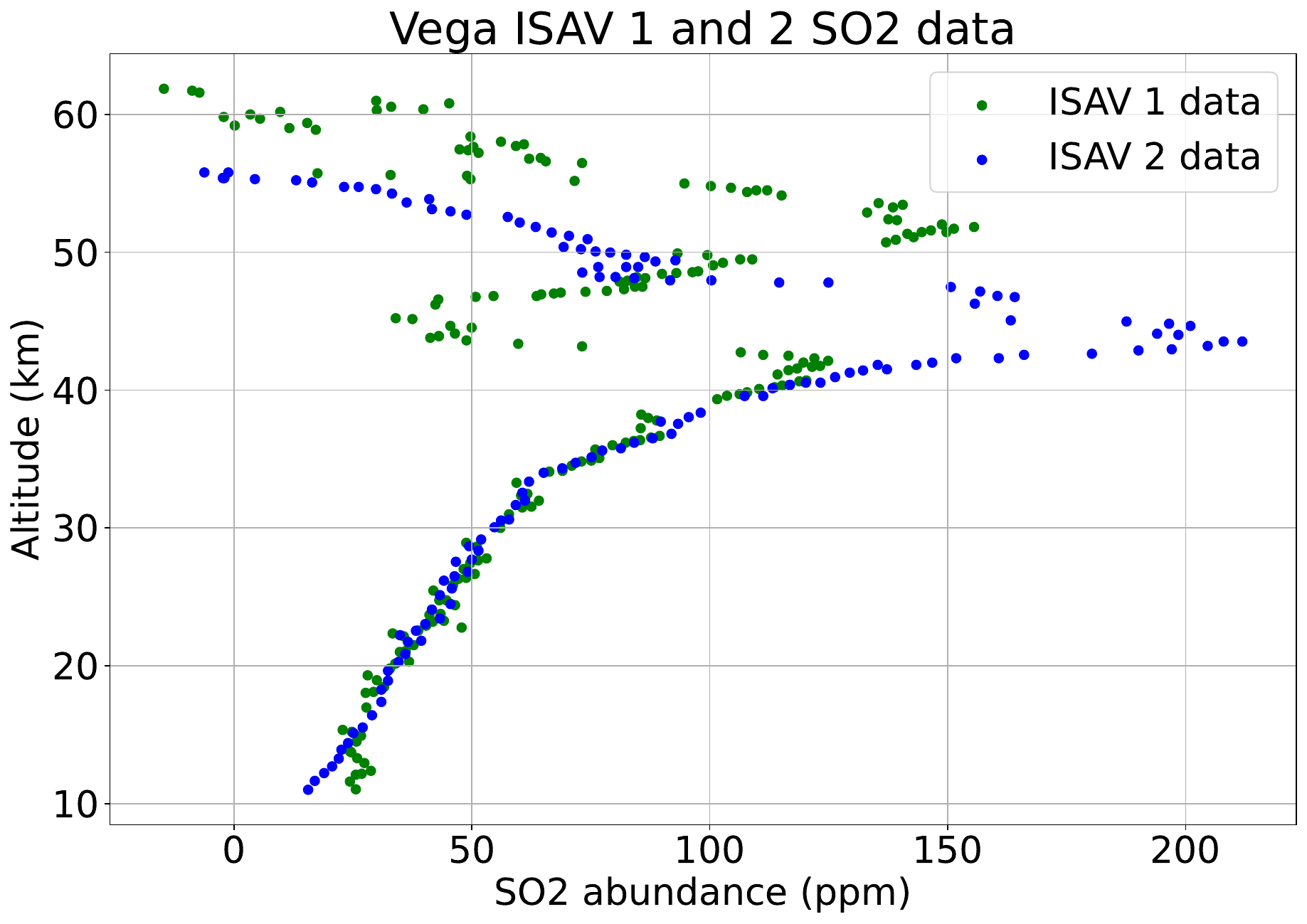} &
\includegraphics[scale=0.25]{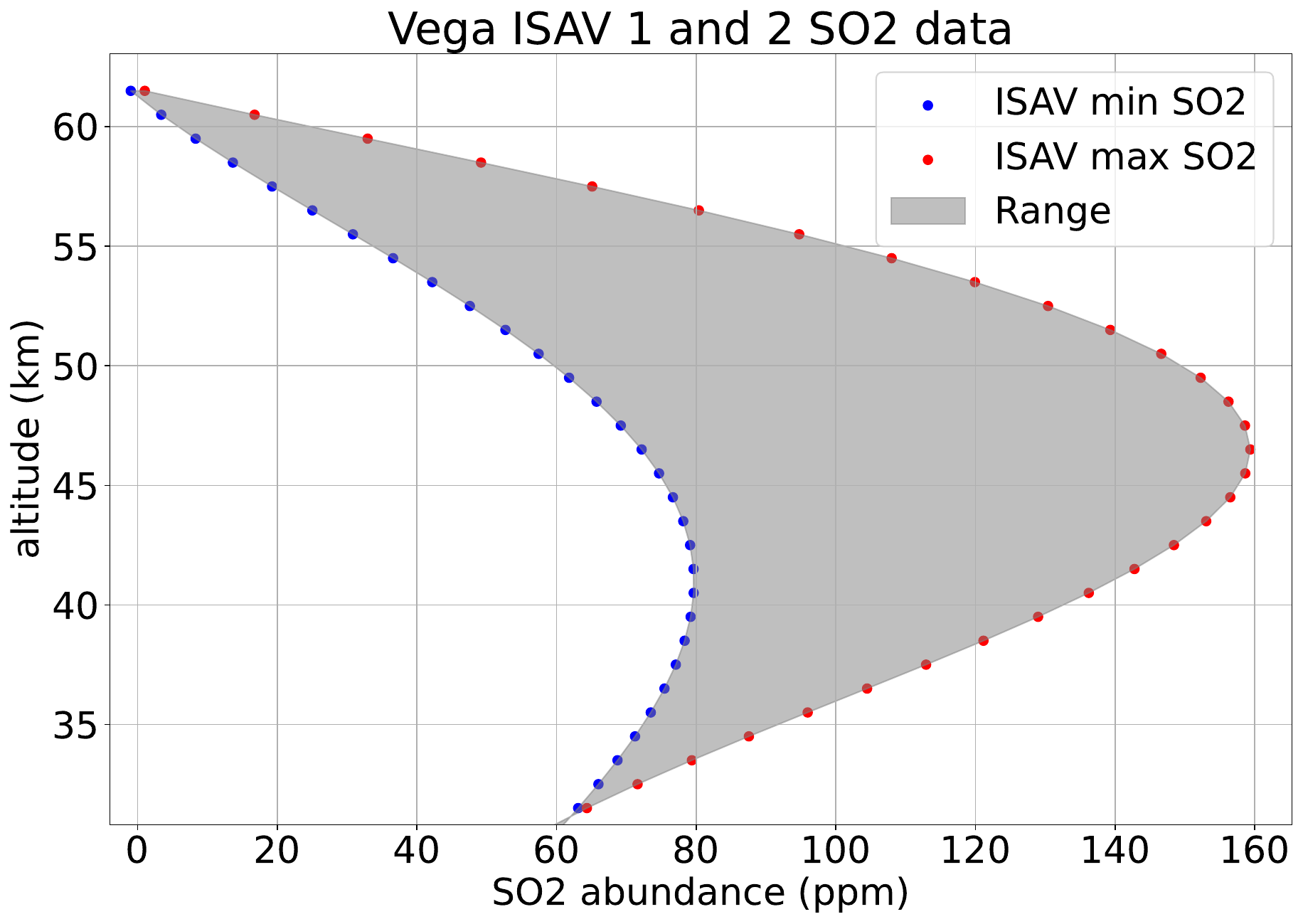} \\
\includegraphics[scale=0.25]{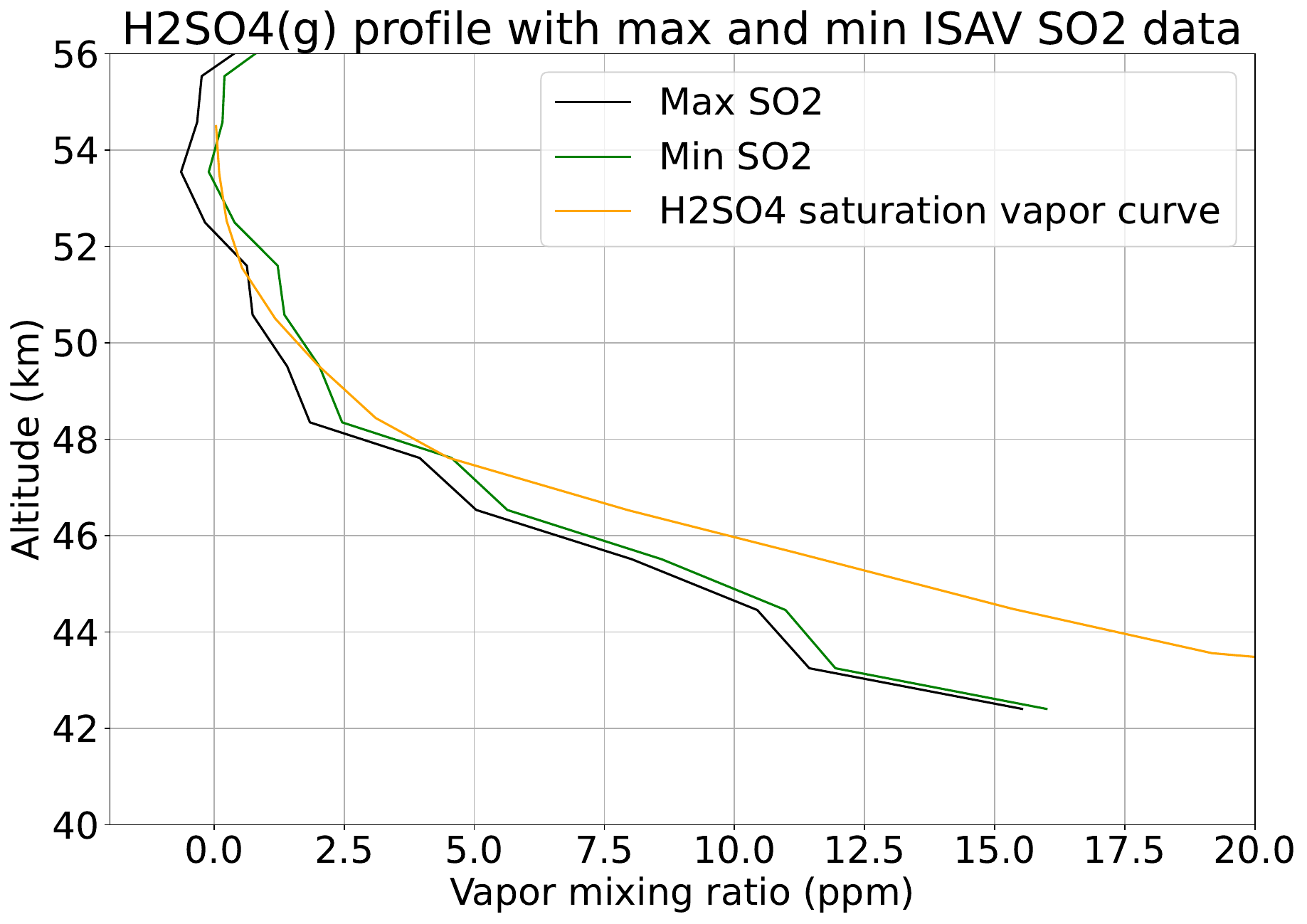} &
\includegraphics[scale=0.25]{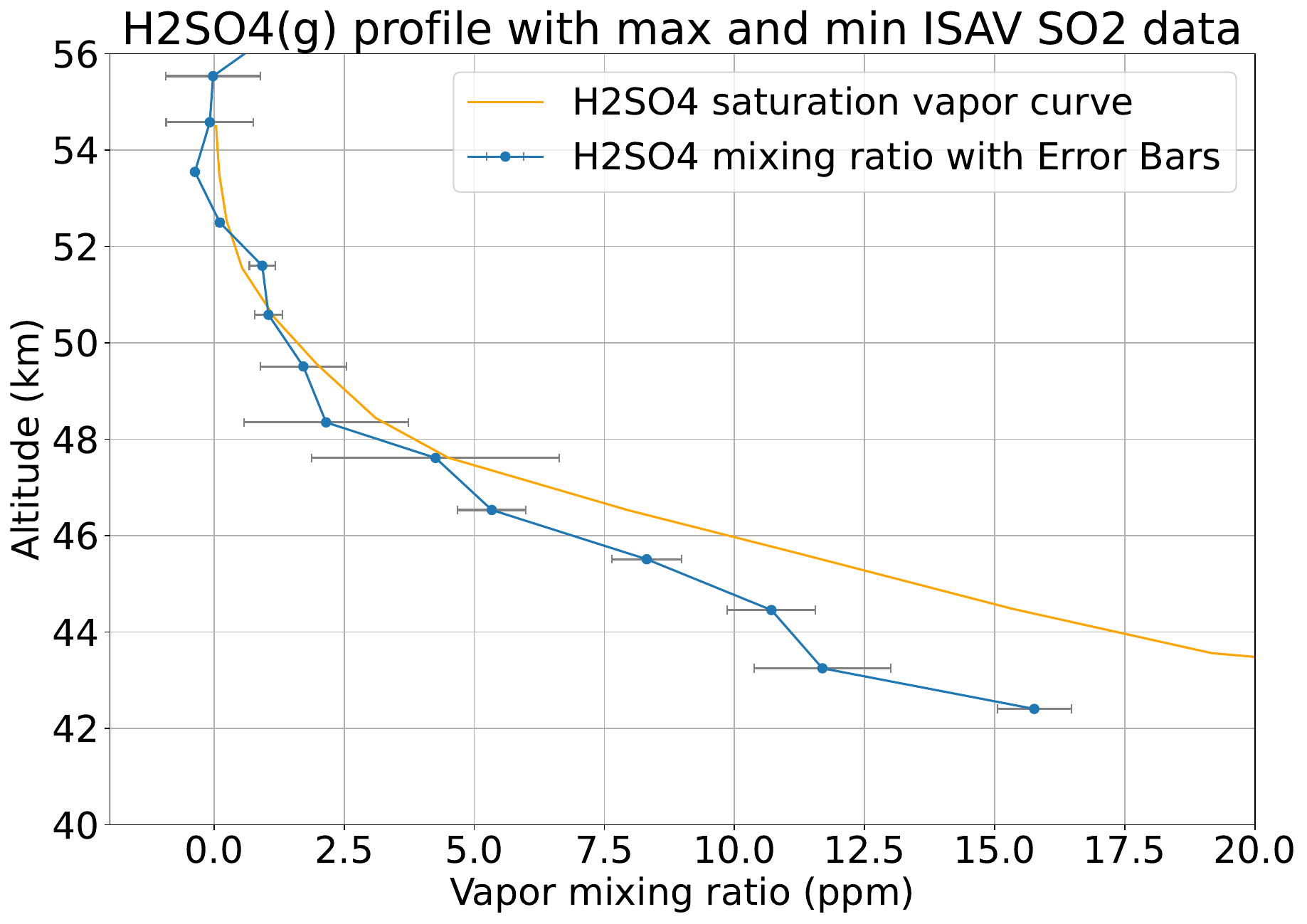}
\end{tabular}
\caption{Estimation of H$_2$SO$_4$ vapor profile with SO$_2$ correction taken from ISAV data. The top left panel shows the in-situ SO$_2$ profiles from the ISAV payloads associated with the VEGA missions \citep{bertaux1996vega}. The green points are from ISAV 1 and the blue ones are from ISAV 2. They show significant deviation from one another above 40 km, and are seen to merge below 40 km. Merged dataset from the two profiles are put into 1 km bins, and two polynomial fits are applied, one with the maximum values and the other with the minimum values from each bin. This is shown in the top right panel, where the red boundary represents the maximum values of the SO$_2$ concentrations and the blue curve represents the minimum values. The H$_2$SO$_4$ vapor profiles are estimated using these maximum and minimum SO$_2$ values as the correction factors to the known absorptivity component. Higher SO$_2$ concentrations leads to lower H$_2$SO$_4$ values. So the black curve in the bottom left panel is the H$_2$SO$_4$ profile for the higher SO$_2$ abundance and the green curve is the H$_2$SO$_4$ profile for the lower SO$_2$ abundance. The orange curve is the saturation vapor abundance. The blue curve in the bottom right panel shows the mean H$_2$SO$_4$ vapor profile calculated by averaging the maximum and minimum H$_2$SO$_4$ abundance values from the bottom left panel. The error bars are calculated using standard error propagation.}
\label{fig:vega}
\end{figure*}

The combined SO$_2$ profile is shown in Figure \ref{fig:vega} (top left) where the green points correspond to the ISAV 1 profile and the blue points are those from ISAV 2. Although the entire dataset covers the altitude range of 10-60 km, for the purposes of our study, we have only considered the data from 30-60 km altitudes. The profiles show considerable deviation from each other. The ISAV 1 profile reveals two local peaks near 50 km (Peak 1) and 40 km (Peak 2) and a trough in between, close to 45 km altitude. The two peaks show vapor abundances of $\sim150$ ppm (for Peak 1) and $\sim120$ ppm (for Peak 2), while the abundance at the trough goes all the way down to $\sim25-30$ ppm. The ISAV 2 shows a single dominant SO$_2$ peak close to 42 km with an abundance of $\sim200$ ppm. The two profiles merge below 35 km and show good agreement with one another all the way down to 15 km. Details about the experiment and the profile features can be found in \citep{bertaux1996vega}.

Despite the variations in the vapor profiles, the peak SO$_2$ abundances of 100-200 ppm in the 40-50 km region from both the datasets roughly agree with other in-situ and ground-based studies on Venus \citep{gelman1979gas, pollack1993near, bezard1993abundance}. An excellent compilation of all the reported vertical distribution of SO$_2$ vmr on Venus obtained by various instruments across four decades (1979-2017) is presented in the Ph.D. Thesis of Daria Evdokimova \citep{evdokimova:tel-03230093}.

Since these VEGA profiles are the only available vertical profiles of Sulfur Dioxide vapor for the Venusian atmosphere in the 35-55 km altitude range, we decided to incorporate all their data points to set a reference for the possible vapor abundance of SO$_2$. The ISAV 1 and 2 profiles are combined to form one dataset, which represents all the plausible SO$_2$ vmr values based on the in-situ data.

To simplify the analysis, the data points are put into 1 km bins, and the maximum and the minimum values corresponding to each bin are recorded. This results in two sets of vertical profiles  - one with the maximum and the other with the minimum SO$_2$ abundance with an altitude resolution of 1 km. These sets are isolated and fifth order polynomial fits are applied to capture the trends in each. These fits then provide us with a range of SO$_2$ abundance values for each altitude, as depicted in Figure \ref{fig:vega} (top right) by the gray region between the maximum (red) and the minimum (blue) profiles. As discussed earlier, the two ISAV profiles coincide below 35 km, and that can be seen with the merging of the red and the blue trends as well.

The X-band absorptivity $\alpha_{SO_2}$ due to SO$_2$ vapors is estimated using Equation \ref{eq:alpha_SO2} with the extrema values. This $\alpha_{SO_2}$ is then used to update the $\alpha_{known}$ factor given in Section \ref{subsec:absorption_loss} and the corresponding $\alpha_{H_2SO_4}$ values are calculated. Finally, the H$_2$SO$_4$ vapor abundance is estimated for the maximum and the minimum sets of SO$_2$ values (Figure \ref{fig:vega} bottom left panel). The mean of the two profiles is plotted in Figure \ref{fig:vega} (bottom right) with the propagated errors. It can be observed that even though a significant difference in the SO$_2$ vapor abundance exists, it does not translate to a big change in the sulfuric acid vapor profile, since Sulfur Dioxide is itself a minor X-band absorber.

\subsection{Indirect Approach}
\begin{figure*}[tbh]
\centering
\begin{tabular}{cc}
\includegraphics[scale=0.25]{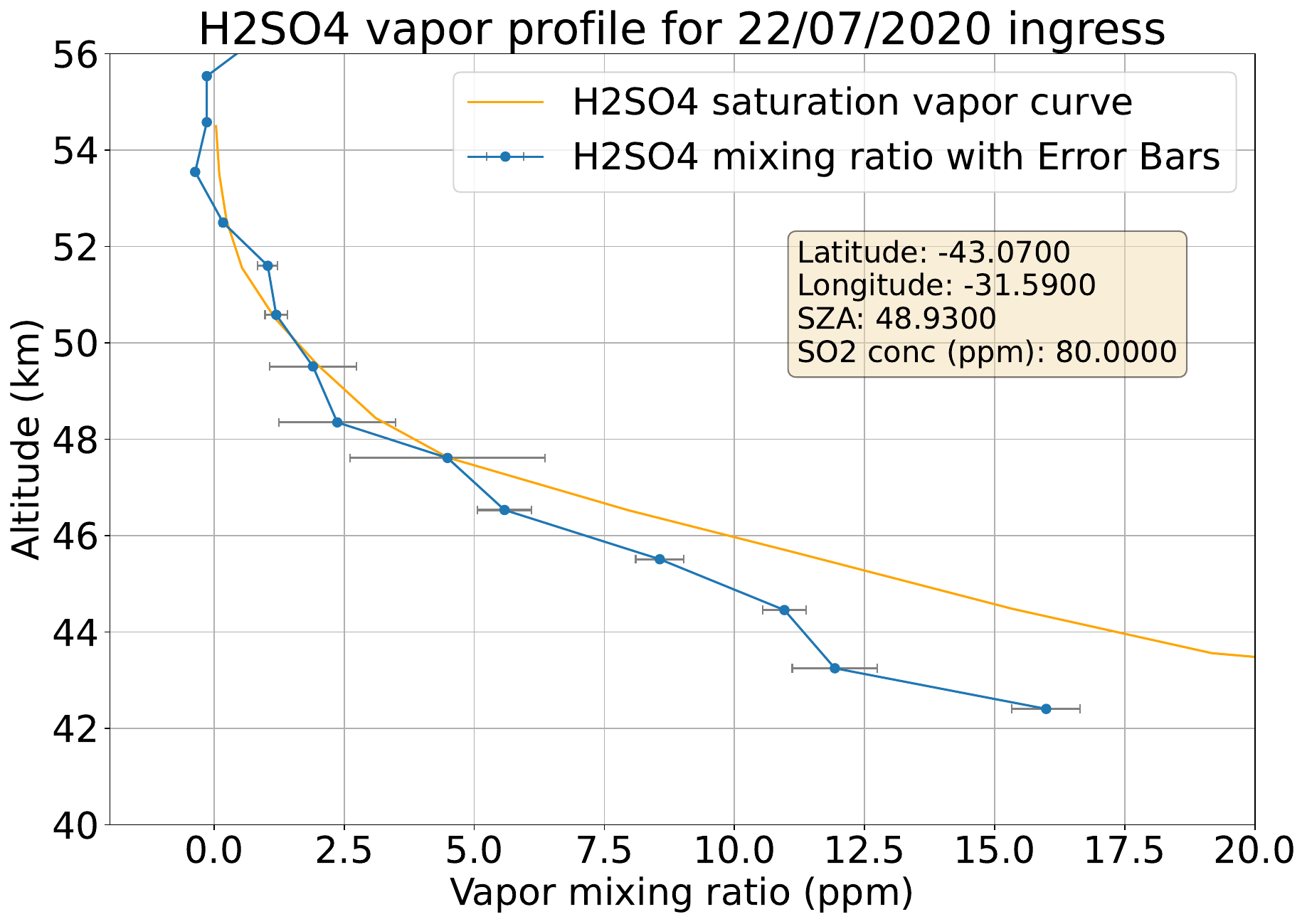} &
\includegraphics[scale=0.25]{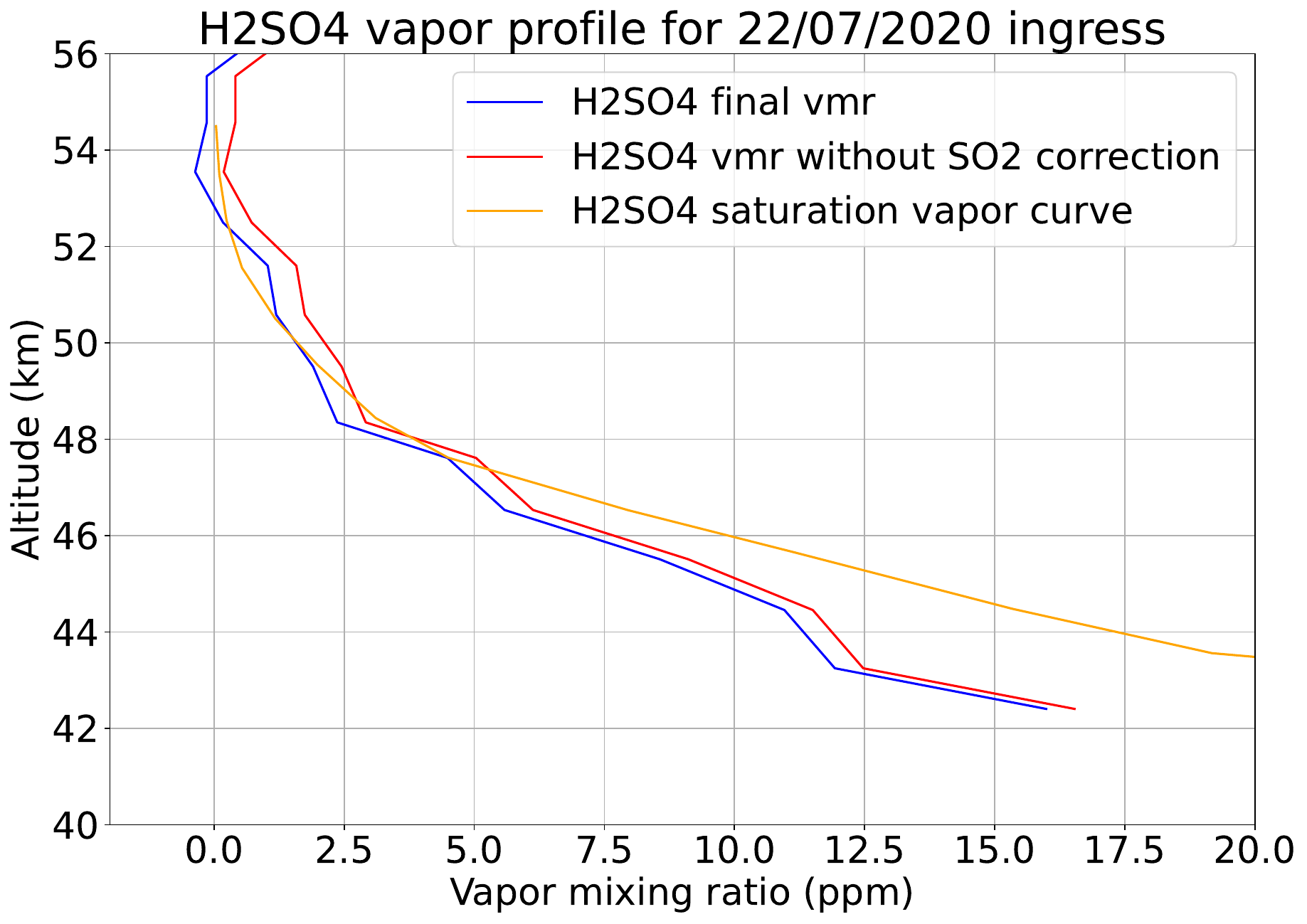}

\end{tabular}
\caption{H$_2$SO$_4$ vapor profile and SO$_2$ vmr estimation for 22 July 2020 ingress ($\sim -43 \degree$ latitude) using the method of \citep{oschlisniok2021sulfuric}. The final H$_2$SO$_4$ vapor profile is shown in the left panel (blue curve). The errors are estimated using standard error propagation. The orange curve is the saturation vapor abundance. In the right panel, the red curve represents the estimated H$_2$SO$_4$ vapor profile assuming null SO$_2$ abundance. The SO$_2$ concentrations are then increased iteratively until the difference between the saturation vapor curve and the H$_2$SO$_4$ vapor abundance in the 51-54 km altitude range is minimum, shown by the blue curve. }
\label{fig:sulfuric}
\end{figure*}

This retrieval technique, first proposed by \citep{oschlisniok2021sulfuric}, begins by assuming a null SO$_2$ abundance to compute the H$_2$SO$_4$ vapor profile using the radiative transfer inversion, shown as the red curve in Figure \ref{fig:sulfuric} (right panel). Above $\sim$50 km, the retrieved H$_2$SO$_4$ profile exhibits non-zero concentrations despite the saturation vapor pressure curve tending asymptotically to zero. The underlying assumption is that this apparent supersaturation arises due to underestimated SO$_2$ opacity.

An iterative inversion loop is employed whereby SO$_2$ vmr is incrementally increased and the H$_2$SO$_4$ profile recalculated at each step. A least-squares cost function is minimized over the 51 -54 km altitude range to identify the SO$_2$ vmr that yields the best agreement between the retrieved and saturation H$_2$SO$_4$ profiles. This optimized vmr is then held constant between 35 -50 km, consistent with prior convention, and incorporated into the X-band absorption coefficient using Equation \ref{eq:alpha_SO2} to update the final H$_2$SO$_4$ profile.

\begin{figure*}
    \centering
    \includegraphics[scale=0.35]{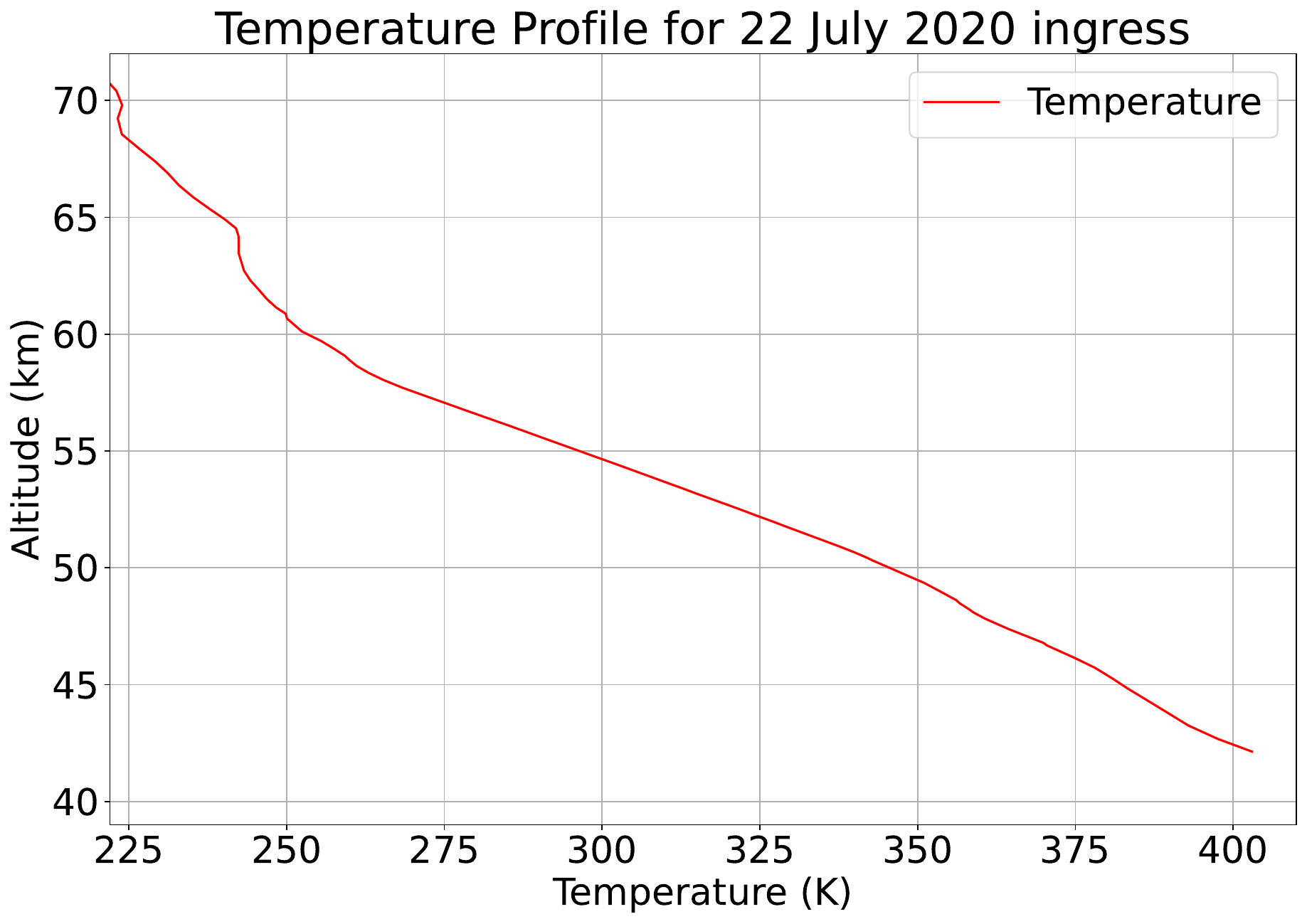}
    \caption{Temperature Profile for 22 July 2020 ingress}
    \label{fig:temperature}
\end{figure*}

Applying this method to the 22 July 2020 ingress occultation yields an SO$_2$ vmr of 80 ppm. The final sulfuric acid vapor profile is shown in Figure \ref{fig:sulfuric} (left panel). The corresponding temperature profile for the given day of the experiment is also plotted in Figure \ref{fig:temperature}. 

\section{Error Analysis}
\label{sec: Error}
The errors in the vapor abundance profile have been estimated using standard error propagation method, discussed extensively in previous studies \citep{lipa1979statistical, jenkins1991results, oschlisniok2012microwave}. A brief account has been presented here. The initial data products are (a) the power of the received signal P, and (b) the frequency residual $f_r$. The uncertainty in P, $\Delta P$, is estimated by taking a 3 point smoothing of the P value and subtracting the original signal power from the averaged data. The uncertainty in $f_r$ however, is very small, primarily because of the source of the radio signal being a USO onboard the Akatsuki orbiter, and is neglected. An extensive error budget has been presented in \citep{tripathi2022quantification} for the Akatsuki RS experiment, where the total uncertainty in the frequency residual is shown to be typically of the order of a few mHz.

Next, for the derived data products, like the impact parameter $a$, the bending angle $\delta$, the ray periapsis $r$, pressure $p$ and temperature $T$, the uncertainties are calculated using the same way as that for the signal power $P$. The refractive index of the atmospheric medium is derived using an Abel inversion as described in Section \ref{subsec:phase}. 

For a function y(x), considering the errors in x to be small, the uncertainty in y can be estimated by applying Taylor series expansion and taking only the linear terms,
\begin{equation}
    \Delta y = \sum_{i=1}^{N} \frac{\partial y}{\partial x_i} \Delta x_i
    \label{eq:Taylor}
\end{equation}

where N is the total number of variables in $x$ with uncertainty $\Delta x$.
The above formalism is applied to  estimate the uncertainty in the refractive index of the medium. Equation \ref{eq:abel_refractive} is linearized about zero, the integral is numerically calculated using the Gaussian quadrature method, and the result is linearized with respect to $\delta$ and $a$ following \citep{jenkins1991results}, such that,
\begin{equation}
    \Delta \mu = \frac{\partial \mu}{\partial a} \Delta a + \frac{\partial \mu}{\partial \delta} \Delta \delta
\end{equation}

Next, the uncertainty in the excess attenuation $\tau$ is be expressed as,
 \begin{equation}
     \Delta \tau = \frac{\partial \tau}{\partial P} \Delta P + \frac{\partial \tau}{\partial a} \Delta a + \frac{\partial \tau}{\partial \delta} \Delta \delta
 \end{equation}

\noindent $\tau$ is converted into absorptivity values $\alpha$ using an Abel inversion as shown in Equation \ref{eq:alpha_abel}, and this inversion is mediated by an integral, termed as F in previous works \citep{jenkins1991results, oschlisniok2012microwave}, given by,
\begin{equation}
    F(\tau, a) = \int_{a(r_0)}^{\infty} \frac{\tau(a) \cdot a    da}{\sqrt{a^2 - a(r_0)^2}}
\end{equation}

Since F is a function of $\tau$ and $a$ only, its uncertainty $\Delta F$ is written as,
\begin{equation}
    \Delta F = \frac{\partial F}{\partial \tau} \Delta \tau + \frac{\partial F}{\partial a} \Delta a 
\end{equation}

After finding $\Delta F$, we then proceed to estimate the uncertainty in the total X-band absorptivity $\alpha$, given by,
\begin{equation}
    \Delta \alpha = \frac{\partial \alpha}{\partial F} \Delta F + \frac{\partial \alpha}{\partial a} \Delta a + \frac{\partial \alpha}{\partial \mu} \Delta \mu
\end{equation}

The uncertainty in the $\alpha$ values of the minor absorbers can be calculated by applying Equation \ref{eq:Taylor} to Equations \ref{eq:alpha_CO2} and \ref{eq:alpha_SO2},

\begin{equation}
    \Delta \alpha_{CO_{2},N_{2}} = \frac{\partial \alpha_{CO_{2},N_{2}}}{\partial T} \Delta T + \frac{\partial \alpha_{CO_{2},N_{2}}}{\partial p} \Delta p
    \label{eq:delta_CO2}
\end{equation}

\begin{equation}
    \Delta \alpha_{SO_{2}} = \frac{\partial \alpha_{SO_{2}}}{\partial T} \Delta T + \frac{\partial \alpha_{SO_{2}}}{\partial p} \Delta p
    \label{eq:delta_SO2}
\end{equation}
to give the uncertainty in the absorptivity of the sulfuric acid vapors,

\begin{equation}
    \begin{split}
        \Delta \alpha_{H_{2}SO_{4}} & = \frac{\partial \alpha_{H_{2}SO_{4}}}{\partial \alpha} \Delta \alpha + \frac{\partial \alpha_{H_{2}SO_{4}}}{\partial \alpha_{CO_{2},N{2}}} \Delta \alpha_{CO_{2},N_{2}} + \\
        & \qquad \frac{\partial \alpha_{H_{2}SO_{4}}}{\partial \alpha_{SO{2}}} \Delta \alpha_{SO_{2}}
    \end{split}
\end{equation}

    

The uncertainty in the H$_2$SO$_4$ vapor abundance can be expressed as,

\begin{equation}
    \begin{split}
        \Delta q_{H_{2}SO_{4}} & = \frac{\partial q_{H_{2}SO_{4}}}{\partial \alpha _{H_{2}SO_{4}}} \Delta \alpha _{H_{2}SO_{4}} + \frac{\partial q_{H_{2}SO_{4}}}{\partial p} \Delta p + \\
        & \qquad \frac{\partial q_{H_{2}SO_{4}}}{\partial T} \Delta T
    \end{split}
\end{equation}

In the region of the Venusian atmosphere considered in the study, $\Delta T \approx 0.1 k$ and $\Delta P \approx$ few hundred Pa are very small, and hence the uncertainty in $q_{H_{2}SO_{4}}$ have been taken to be independent of the $\Delta T$ and the $\Delta P$ values. Finally, we can calculate the uncertainty in the abundance of the Sulfuric acid vapor abundance as,
\begin{equation}
    \Delta q_{H_{2}SO_{4}} = \frac{\partial q_{H_{2}SO_{4}}}{\partial \alpha _{H_{2}SO_{4}}} \Delta \alpha _{H_{2}SO_{4}}
\end{equation}

The error bars in the H$_2$SO$_4$ vapor profiles in Figures  \ref{fig:vega} and  \ref{fig:sulfuric} have been estimated from the above equation.

\subsection{Choice of Polynomial fit for the residuals}
As stated in Section \ref{subsec:calibration_resi}, for the accurate determination of the frequency residuals from the noisy regions of the atmosphere, polynomial fits of various orders are applied on the non-fluctuating residuals corresponding to the regions in the neutral Venusian atmosphere below 100 km. These trends are then extrapolated forward/backward in time to probe deeper into the atmosphere during ingress/egress. The rms error is then calculated for multiple trends (on comparison with the actual residuals), and the one with the least error is chosen as the preferred fit for that part (ingress/egress) of the occultation experiment.

\begin{figure*}[tbh]
\centering
\begin{tabular}{cc}
\includegraphics[scale=0.25]{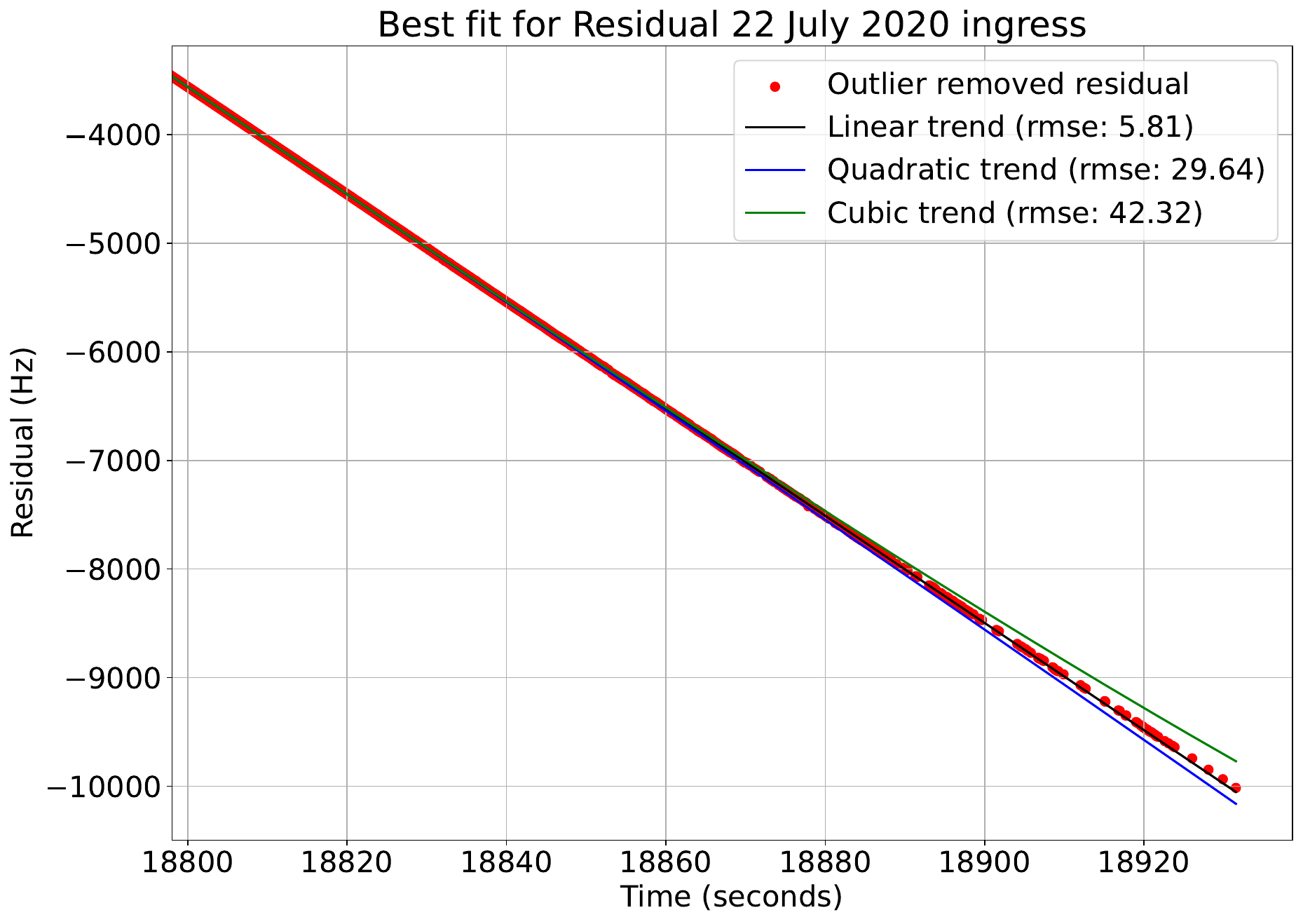} &
\includegraphics[scale=0.25]{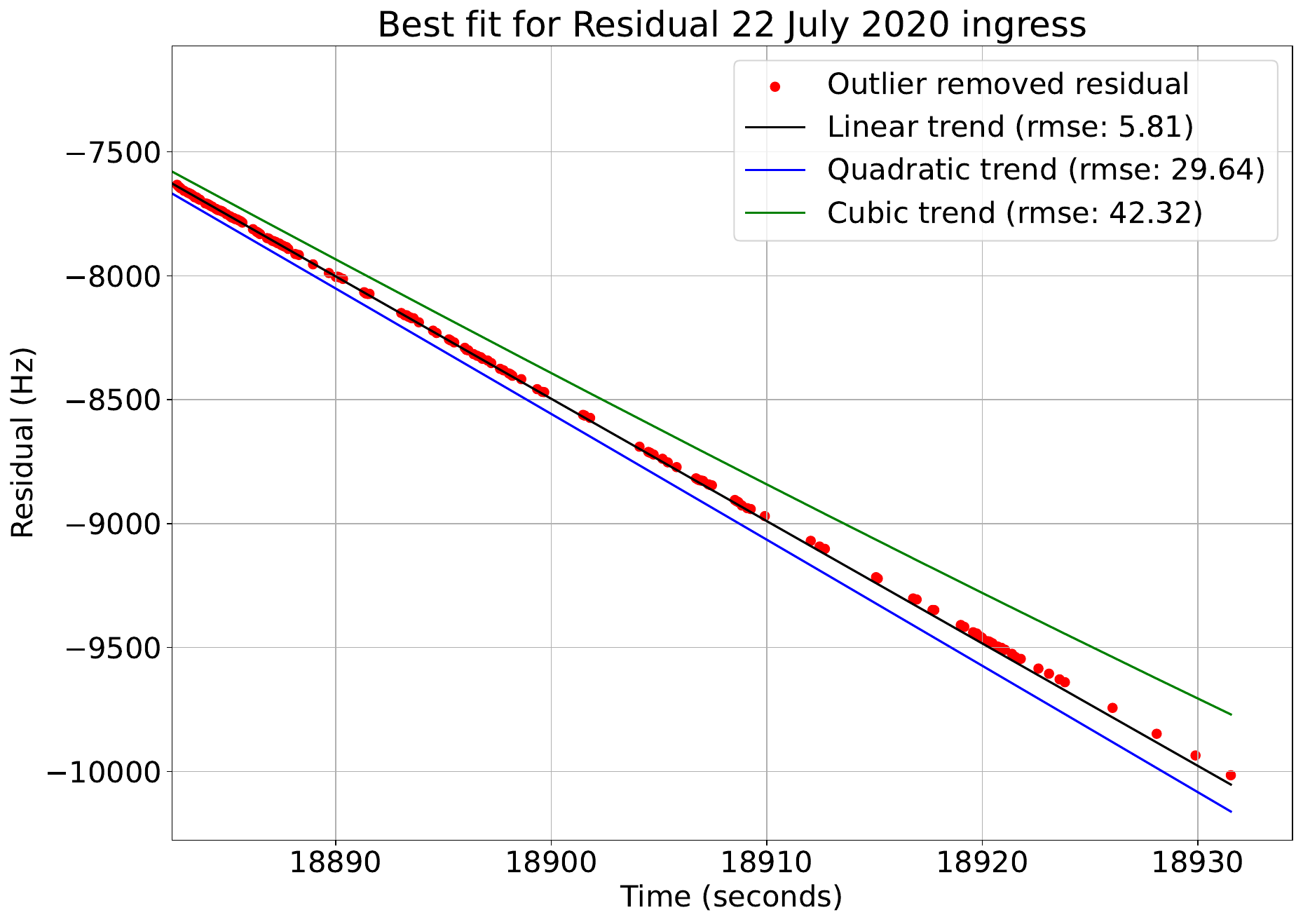} \\
\includegraphics[scale=0.25]{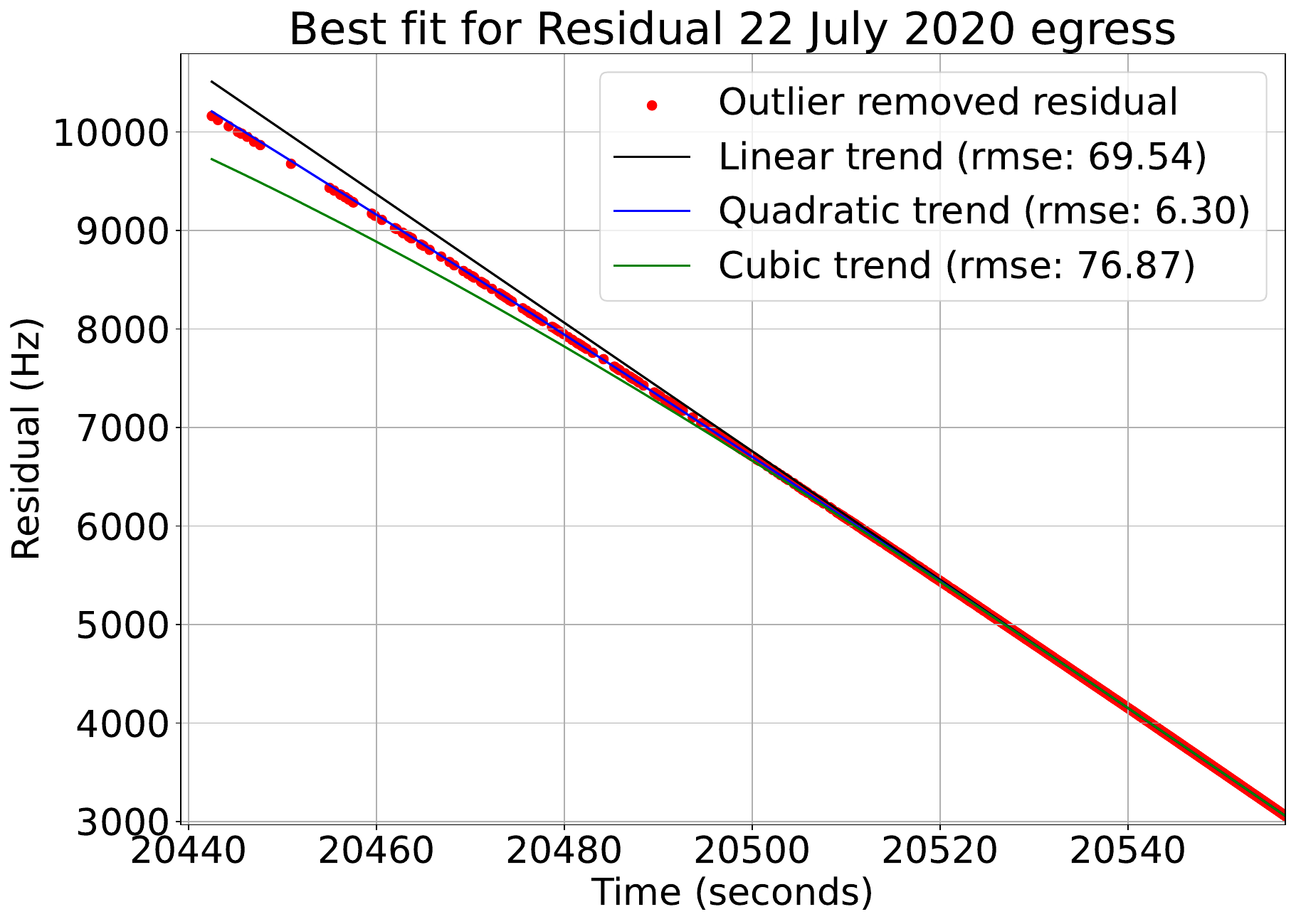} &
\includegraphics[scale=0.25]{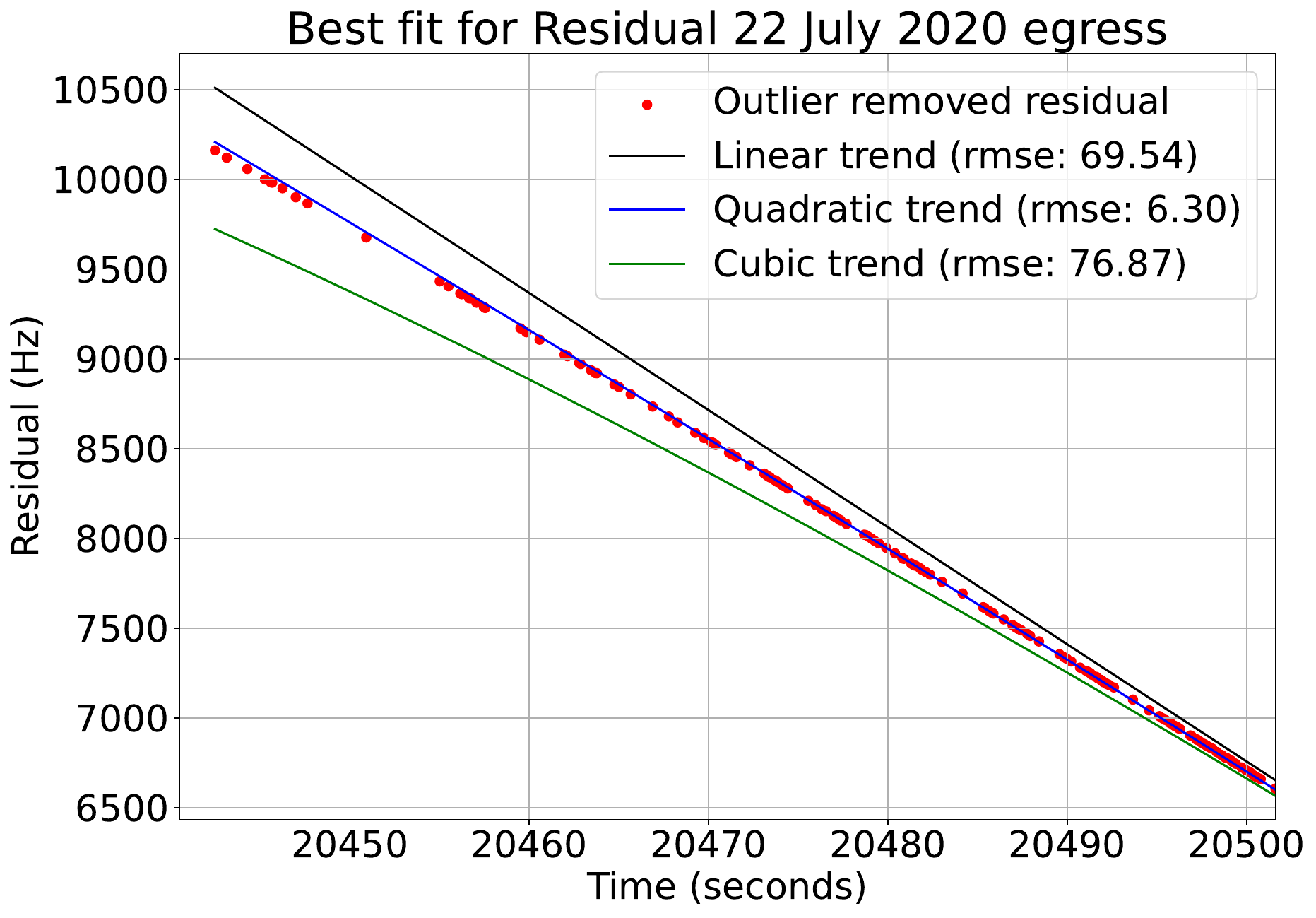}
\end{tabular}
\caption{Polynomial fits to capture frequency residual trends for 22 July 2020. The top panels show the residuals for the ingress phase of the experiment, while the bottom panels represent the egress part . The top right and the bottom right panels are the respective zoomed in profiles of the top left and bottom left ones at the deepest altitudes, where the clear deviations among the different trends are observed. The rms error is estimated for each trend on comparing it with the residuals (red scatter points). It is observed that for the ingress phase, a linear trend shows the least error, while for the egress phase, the residual trend is captured best by a quadratic fit.}
\label{fig:rmse}
\end{figure*}

For the 22 July 2020 RO experiment, we find that a linear trend in the residuals for the ingress phase shows the least error and is able to capture the trend quite well (Figure \ref{fig:rmse} top panels), while for the egress phase the second order polynomial trend shows the highest accuracy (Figure \ref{fig:rmse} bottom panels).

\section{Concluding remarks}

The formation and structure of the sulfuric acid vapor layer beneath the Venusian cloud deck are critical to constraining meridional circulation patterns, including Hadley-like cells, within the middle atmosphere. This layer’s upper boundary coincides with the cloud-vapor transition region, where thermal gradients are expected to modulate vapor saturation. Among available remote sensing methods, radio occultation (RO) uniquely enables routine, vertically resolved monitoring of this region across multiple occultation seasons.

In this study, we have introduced a signal processing algorithm for deriving H$_2$SO$_4$ vapor profiles from X-band RO measurements. The power spectrum of the raw received signal is analyzed. The statistical moments of the power spectral density (PSD) provides the total received power and the Doppler shift in the signal. Signal from the deep Venusian noisy atmosphere is extracted. The frequency residuals are averaged across the First Fresnel Zone (FFZ). These FFZ averaged residuals are then used to estimate the defocusing loss in the signal, as well as other neutral atmospheric parameters, such as the temperature and the pressure of the medium.  The total signal attenuation increases below 50 km, primarily due to the Sulfuric acid vapors present in the region. The absorption loss is then converted into specific attenuation values, from which the loss due to known components are removed, leaving behind the absorption due to the sulfuric acid vapors. 

To get an accurate estimate of the H$_2$SO$_4$ vapor abundance, it is necessary to characterize the Sulfur dioxide vapor concentration, which is one of the inputs to the known component of the signal absorption loss. A couple of approaches are used to approximate this SO$_2$ abundance  - one by directly taking the altitudinally averaged profiles given by the ISAV payloads onboard the VEGA spacecrafts, and the other by implementing the iterative retrieval scheme of \citet{oschlisniok2021sulfuric}, the latter yielding SO$_2$ vmr of 80 ppm, with values from both approaches roughly agreeing with the expected range for the sub-cloud region. The final retrieved H$_2$SO$_4$ vapor profiles from both methods show similar trends, the abundance increasing with depth and reaching a peak value exceeding 15 ppm below 43 km. Above 50 km, the SO$_2$ inputs from the VEGA data as well as the \citet{oschlisniok2021sulfuric} method result in near-zero H$_2$SO$_4$ vapor concentrations.

Ongoing analysis of the complete IDSN RO dataset from the Akatsuki mission will further refine these results. The proposed retrieval framework is also applicable to future dual-frequency (X/Ka-band) RO missions, including ISRO’s planned Venus Orbiter Mission, which may enable simultaneous probing of vapor and cloud layers in the Venusian middle atmosphere.

\begin{acknowledgments}
We sincerely acknowledge the help and support of the ground station team at IDSN, India to track the Akatsuki spacecraft. Special thanks to Umang Parikh, Anshuman Sharma, D32 Antenna, IDSN; and Himanshu Pandey, ISSDC, at Byalalu, India, for their proactive support. We acknowledge the support of Simi RS for archiving the data at SPL and making them available for analysis. We also thank Dr. Vijayakumar S. Nair at SPL for fruitful discussions on the estimation of the theoretical saturation vapor pressure on Venus.
\end{acknowledgments}

\begin{contribution}
SB was responsible for analyzing the data, writing and submitting the manuscript. RKC came up with the initial research concept, supervised the project and edited the manuscript. KRT, and HA helped the formal analysis and validation. They also helped in editing the manuscript. TI provided the IDSN data and helped in editing the manuscript.

\end{contribution}

\facilities{IDSN}
\software{Python}
\bibliography{Paper1_H2SO4.bib}{}
\bibliographystyle{aasjournalv7}

\end{document}